\documentclass[twocolumn,showpacs,showkeys,preprintnumbers,amsmath,amssymb]{revtex4}

\usepackage{graphicx}
\usepackage{dcolumn}
\usepackage{bm}
\usepackage{multirow}
\usepackage{color}
\usepackage{CJK}
\usepackage{threeparttable} 
\usepackage{booktabs}
\usepackage{longtable}

\newcommand{\br}{\mbox{\boldmath$r$}}
\newcommand{\bj}{\mbox{\boldmath$j$}}
\newcommand{\svec}[1]{{\mbox{\boldmath$#1$}}}
\newcommand{\balp}{\mbox{\boldmath$\alpha$}}
\newcommand{\bp}{{\mathbf{p}}}
\newcommand{\bnab}{\mbox{\boldmath$\nabla$}}
\newcommand{\bmu}{\mbox{\boldmath$\mu$}}
\newcommand{\bSig}{\mbox{\boldmath$\Sigma$}}

\begin{document}

\title{Nuclear magnetic moments in covariant density functional theory}

\author{Jian Li~$^{1,2}$}

\affiliation{$^1$College of Physics, Jilin University, Changchun 130012, China}
\affiliation{$^2$Department of Physics, Western Michigan University, Kalamazoo, MI 49008, USA}

\author{J. Meng$^{3,4,5}$}
 \email{mengj@pku.edu.cn}
 \affiliation{$^3$State Key Laboratory of Nuclear Physics and Technology, School of Physics, Peking University, Beijing 100871, China}
 \affiliation{$^4$Yukawa Institute for Theoretical Physics, Kyoto University, Kyoto 606-8502, Japan}
  \affiliation{$^5$Department of Physics, University of Stellenbosch, Stellenbosch, South Africa}


\begin{abstract}
Nuclear magnetic moment is an important physical observable and serves as a useful tool for the stringent test of nuclear models. For the past decades, the covariant density functional theory
and its extension have been proved to be successful in describing the nuclear ground-states
and excited states properties. However, a long-standing problem is its
failure to predict magnetic moments. This article reviews the recent progress in the description of the  nuclear magnetic moments  within the covariant density functional theory. In particular, the magnetic moments of spherical odd-$A$ nuclei with doubly closed shell core plus or minus one nucleon and deformed odd-$A$ nuclei.
\end{abstract}



\maketitle
\tableofcontents

\section{Introduction}

The longstanding frontiers in nuclear physics include the exploration of fundamental symmetry and the understanding the properties of atomic nucleus. The pseudospin symmetry~\cite{Arima1969Phys.Lett.B517, Meng1998Phys.Rev.CR628,Meng1999Phys.Rev.C154,Liang2015Phys.Rep.1} and magnetic moments~\cite{Arima1987Adv.Nucl.Phys.1,Arima1954Prog.Theor.Phys.509,Horie1955PhysRev.99.778,Shimizu1974Nucl.Phys.A282} are definitely these fascinating examples.
Here in this paper we will concentrate on the description of the magnetic moments in the relativistic approach. In particular, the magnetic moment for nuclei with $LS$ and $jj$ closed shell plus or minus one nucleon~\cite{Li2009Sci.ChinaSer.G1586, Li2011Sci.ChinaPhys.Mech.Astron.204,Li2011Prog.Theor.Phys.1185, Wei2012Prog.Theor.Phys.Supp.400,Li2013Phys.Rev.C064307} will be reviewed.

The magnetic moment is one of the important fundamental properties of the nucleus. It serves as a useful tool for the stringent test of nuclear models, and has attracted the attentions of many nuclear physicists since the early days~\cite{Arima1954Prog.Theor.Phys.509,Horie1955PhysRev.99.778,Blin-Stoyle1956Rev.Mod.Phys.75,
Arima1978Prog.Part.Nucl.Phys.41,
Arima1984Prog.Part.Nucl.Phys.53,
Towner1987Phys.Rep.263,
Castel1990,Talmi2005Int.J.Mod.Phys.E821}.
The theoretical description of the nuclear magnetic moments is a long-standing
problem. Although many successful nuclear structure
models have been developed in the past decades, the application of these
models for nuclear magnetic moments is still not satisfactory.

Since the establishment of the independent particle shell model in 1949 by Mayer and Jensen
to explain the magic numbers $Z$= 2, 8, 20, 28, 50, and 82 as well as $N$ = 2, 8, 20, 28, 50, 82, and 126,
the magnetic moment for an odd-$A$ nucleus has been interpreted as the contribution from the unpaired valence nucleon,
  \begin{eqnarray}\label{Schmidt}
    \mu
       &=&\langle (nl)\,jm|g_l\hat{l}_z+g_s\hat{s}_z|(nl)\,jm\rangle_{m=j} \nonumber \\ \nonumber \\
       &=&  \left\{
             \begin{array}{ll}
               g_ll+\frac{1}{2}g_s, \qquad &\qquad j=l+1/2 \nonumber \\  \nonumber \\
               \dfrac{j}{j+1}[g_l(l+1)-\dfrac{1}{2}g_s],  &\qquad j=l-1/2, \nonumber\\
             \end{array}
           \right.
  \end{eqnarray}
where $j=l\pm1/2$ is the total angular momentum of the valence nucleon, and $g_l=1(0)$ and $g_s=5.59(-3.83)$ are respectively
the orbital and spin $g$-factors of the proton (neutron).  The  magnetic moment in above equation
as a function of spin $j$ will result in the so-called Schmidt lines~\cite{Schmidt1937Z.Phys.A358}.

\begin{figure}[!ht]
\centering
\includegraphics[width=8.5cm]{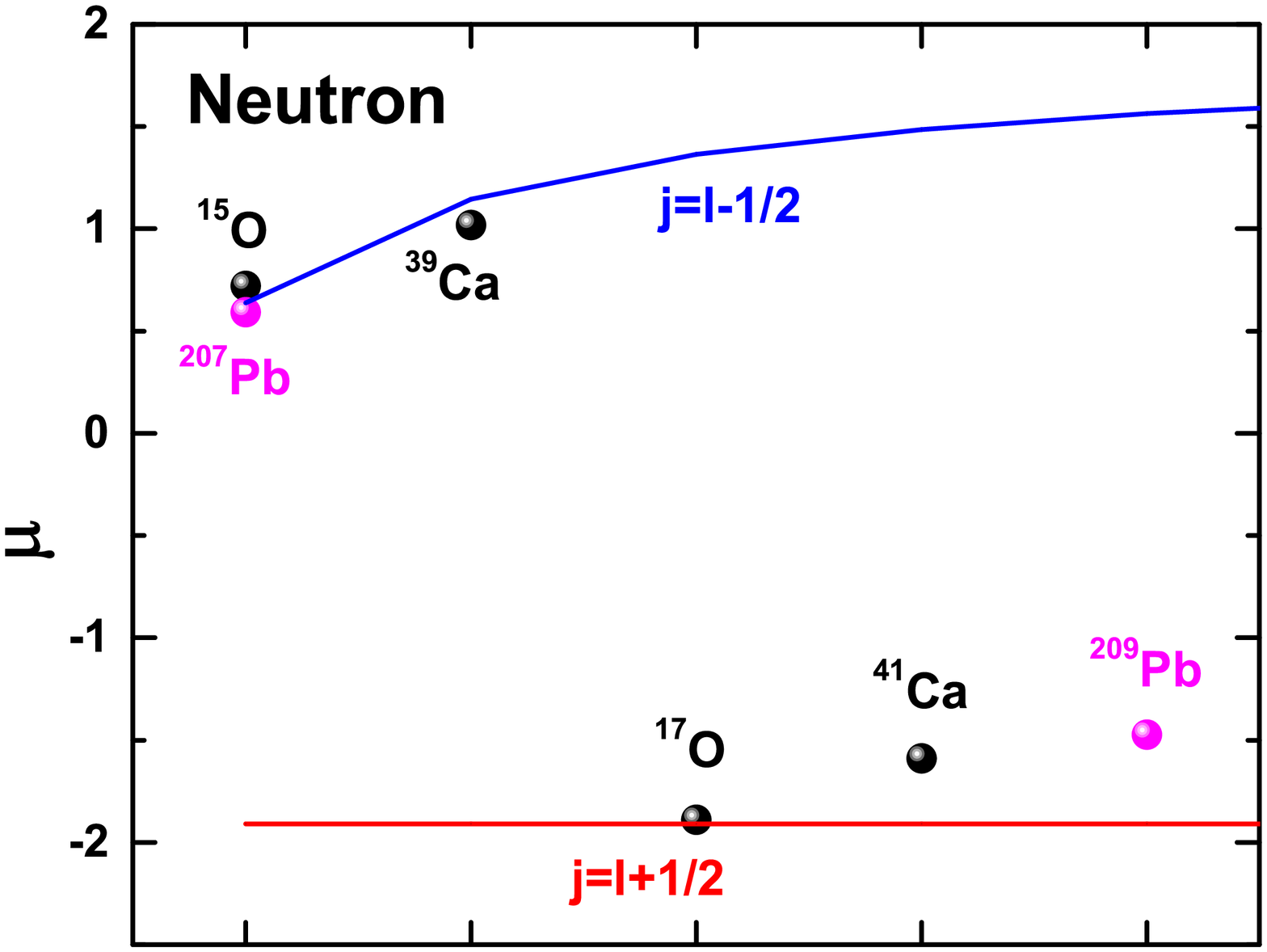}\vspace*{-1.5cm}\\
\includegraphics[width=8.5cm]{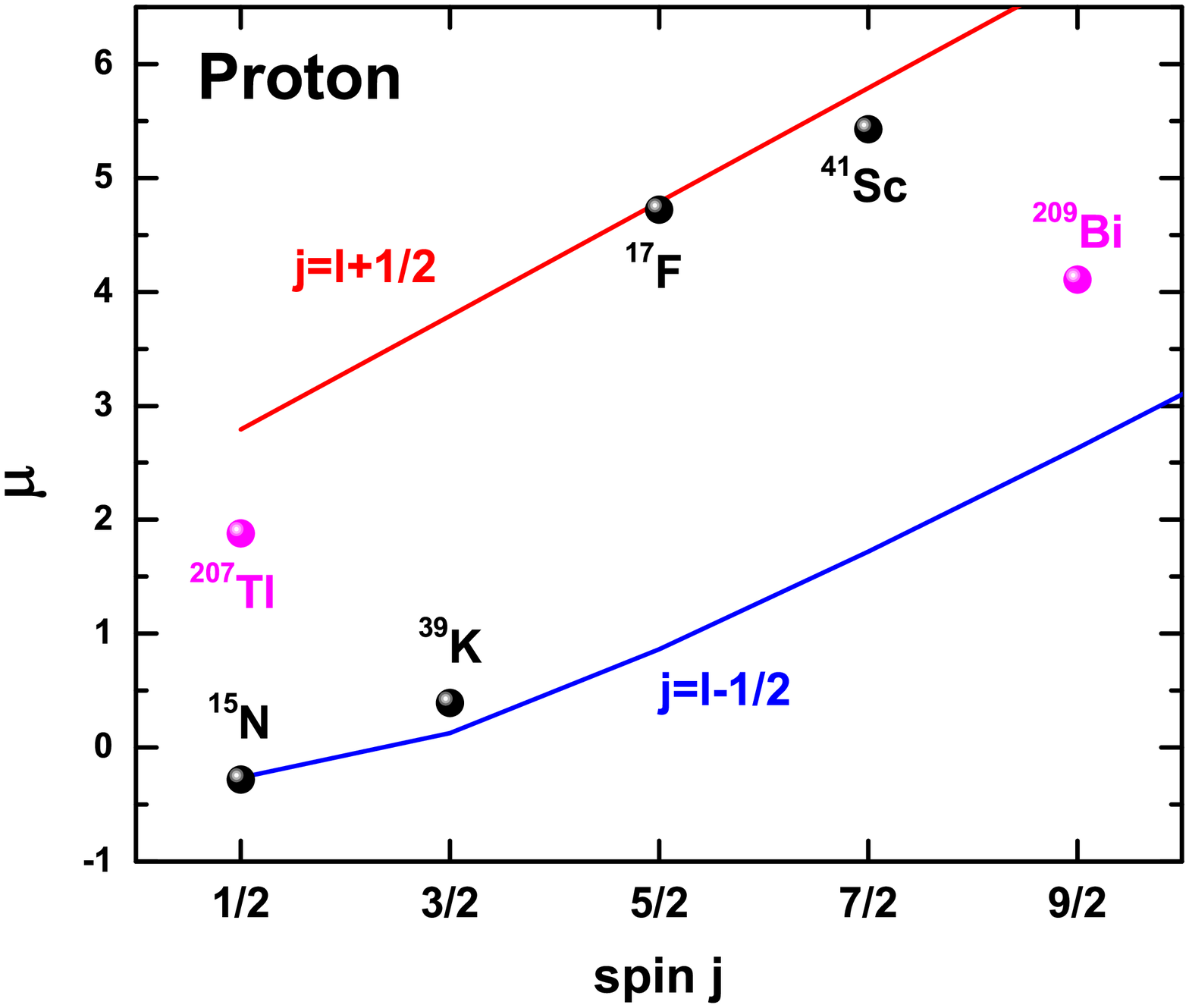}
\caption{Experimental magnetic moments of odd-neutron (upper) and odd-proton (lower) nuclei in the neighbourhood of $^{16}$O, $^{40}$Ca and $^{208}$Pb, in comparison with the corresponding Schmidt values.}
\label{fg-Schmidt}
\end{figure}

It was observed in the early 1950s~\cite{Blin-Stoyle1953Proc.Phys.Soc.A1158,Blin-Stoyle1954Proc.Phys.Soc.A885} that almost all nuclear magnetic moments are sandwiched between the two Schmidt lines. As shown in Fig.~\ref{fg-Schmidt}, some of them are close to the Schmidt lines, like $^{17}$F or $^{15}$N, and some deviate very much, like $^{209}$Bi or $^{207}$Tl. Therefore, a lot of efforts had been made to explain the deviations of the nuclear magnetic moments from the Schmidt values.

In 1954, Arima and Horie~\cite{Arima1954Prog.Theor.Phys.509} pointed out the difference between the following two groups of nuclei. For the first group, the cores are $LS$-closed ($^{16}$O and $^{40}$Ca),
i.e., the spin-orbit partners $j = l \pm \frac{1}{2}$ of the core are completely occupied.
Therefore they are not expected to be
excited by an external field with $M1$ character strongly. For the second group,
the cores are $jj$-closed (like $^{208}$Pb),
i.e., one of the spin-orbit partners is open, and an $M1$ external field can strongly excite the core nucleons to the empty
spin-orbit partner. This $M1$ giant resonance state of the core can be momentarily excited by the interaction with the valence
nucleon. This is the idea of first-order configuration mixing, which is also called the Arima-Horie effect~\cite{Talmi2005Int.J.Mod.Phys.E821}. It explains not only the difference between these two groups of nuclei, but also the deviations of magnetic moments from the Schmidt lines for many nuclei~\cite{Arima1954Prog.Theor.Phys.509,Arima1954Prog.Theor.Phys.623,Hiroshi1958Prog.Theor.Phys.Suppl.33}.

In the 1960s, it became clear that the first-order effect is not enough to explain the large deviations from the Schmidt values in some nucleui, e.g., $^{209}$Bi. The pion exchange current is found to be very important to understand nuclear magnetic moments, as was first pointed out by Miyazawa in 1951~\cite{Miyazawa1951Prog.Theor.Phys.801} and by Villars in 1952~\cite{Villars1952}. The correction changes the gyromagnetic ration of the orbital angular momentum of a nucleon in the nucleus~\cite{Chemtob1969Nucl.Phys.A449} and improves the agreement between theoretical and observed values~\cite{Hyuga1980Nucl.Phys.A363}. Nowadays, the importance of meson exchange currents is a well-established fact.

As mentioned earlier, the first-order configuration mixing does not contribute in nuclei with an $LS$ closed core $\pm1$
nucleon. However, magnetic moments of light nuclei with a valence nucleon or hole outside an $LS$ doubly closed shell have shown appreciable deviations from their single-particle values, including the isoscalar and isovector magnetic moments. The earliest attempts to understand these deviations were carried out in Refs.~\cite{Ichimura1965,Mavromatis1966Nucl.Phys.545,Mavromatis1967Nucl.Phys.A17} by taking into account the second-order configuration mixing as well as the effect of the meson exchange currents.
However, only very small second-order corrections were obtained.
Shimizu, Ichimura and Arima~\cite{Shimizu1974Nucl.Phys.A282} recalculated this effect and paid attention to second-order contributions from intermediate states with higher excitation energies due to the short-range nature of the tensor force. They found that the second-order correction, which is also called the tensor correlation, is obviously needed to explain these deviations. Their results are also in agreement with the later results by Towner and Khanna~\cite{Towner1983Nucl.Phys.A334}.

Considerable efforts have been made to explain the deviations of the nuclear magnetic moments
from the Schmidt values, which can be contributed from the meson exchange current (MEC, i.e.,
the exchange of the charged mesons) and configuration mixing (CM, or core polarization, i.e.,
the correlation not included in the mean field approximation)~\cite{Arima1978Prog.Part.Nucl.Phys.41,Arima1984Prog.Part.Nucl.Phys.53,Towner1987Phys.Rep.263,Arima1987Adv.Nucl.Phys.1,Castel1990,Arima2011Sci.ChinaPhys.Mech.Astron.188}.
Recently, by considering the configuration mixing and meson exchange current corrections, the newly measured magnetic moment of $^{133}$Sb~\cite{Stone1997Phys.Rev.Lett.820}, $^{67}$Ni and $^{69}$Cu~\cite{Rikovska2000Phys.Rev.Lett.1392}, and $^{49}$Sc~\cite{Ohtsubo2012Phys.Rev.Lett.032504} have been well reproduced.

In the past decades, the covariant density functional theory (CDFT), taking Lorentz symmetry into account
in a self-consistent way, has received wide attention due to
its successful description of a large number of nuclear phenomena in stable as well as exotic nuclei~\cite{Ring1996Prog.Part.Nucl.Phys.193,Vretenar2005Phys.Rep.101,Meng2006Prog.Part.Nucl.Phys.470,Meng2011Prog.Phys.199,Niksic2011Prog.Part.Nucl.Phys.519,Meng2015}. It includes naturally the nucleonic spin degree of
freedom and automatically results in the nuclear spin-orbit potential with the empirical strength in a covariant way. It can reproduce
well the isotopic shifts in the Pb region~\cite{Sharma1993Phys.Lett.B9}, and give naturally the
origin of the pseudospin symmetry~\cite{Liang2015Phys.Rep.1} and the spin symmetry
in the anti-nucleon spectrum~\cite{Zhou2003Phys.Rev.Lett.262501}. Furthermore, it can include
the nuclear magnetism~\cite{Koepf1989Nucl.Phys.A61}, that is, a consistent description of
currents and time-odd fields, which plays an important role in
nuclear rotations~\cite{Konig1993Phys.Rev.Lett.3079,Afanasjev2000Phys.Rev.C62.031302,Afanasjev2010PhysRevC.034329,Zhao2011Phys.Lett.B181,Zhao2011Phys.Rev.Lett.122501,Meng2013}.

However, relativistic descriptions of nuclear magnetic moments are mostly restricted to nuclei with $LS$ closed-shell $\pm1$ nucleon. It is known that a straightforward application of the relativistic model, where only the sigma and the time-like component of the vector mesons were considered, the predicted isoscalar magnetic moments were significantly larger than the observed
values~\cite{Miller1975Ann.Phys.40,Serot1981Phys.Lett.B263}.
After introducing the vertex corrections by the effective single-particle currents in nuclei, i.e., the ¡°back-flow¡± effect in the framework of a relativistic extension of Landau's Fermi-liquid theory~\cite{McNeil1986Phys.Rev.C746}, or by a random phase approximation (RPA) type summation of p-h and p-$\bar{n}$ bubbles~\cite{Ichii1987Phys.Lett.B11,Shepard1988Phys.Rev.C1130}, or by the non-zero space-like component of vector mesons~\cite{Hofmann1988Phys.Lett.B307,Furnstahl1989Phys.Rev.C1398,Yao2006Phys.Rev.C024307,Li2009Sci.ChinaSer.G1586}, good agreement with Schmidt magnetic moments can be obtained and the isoscalar magnetic moments for odd-$A$ nuclei in the vicinity of $LS$ closed shells can be reproduced rather well.

Unfortunately, these effects cannot remove the discrepancy existing
in isovector magnetic moments. To eliminate
this discrepancy, one-pion exchange current corrections was found to be significant
in the relativistic model~\cite{Morse1990Phys.Lett.B241,Li2011Prog.Theor.Phys.1185}.
However one-pion exchange current corrections lead to
large disagreement with data. The introduction of the second-order configuration mixing in the fully
self-consistent relativistic theory greatly improves the description of isovector magnetic moments~\cite{Li2011Sci.ChinaPhys.Mech.Astron.204,Wei2012Prog.Theor.Phys.Supp.400}.

In Ref.~\cite{Furnstahl1987Nucl.Phys.A539}, the magnetic moments for nuclei with $jj$ closed-shell $\pm1$ nucleon near $^{208}$Pb have been studied in CDFT including the contribution from the core. The corresponding results show an improvement in comparison with the valence-nucleon approximation. By considering the meson exchange currents and configuration mixing, especially the first-order configuration mixing, the magnetic moments  for nuclei with $jj$ closed-shell $\pm1$ nucleon, $^{207}$Pb, $^{209}$Pb, $^{207}$Tl, and $^{209}$Bi, have been well reproduced~\cite{Li2013Phys.Rev.C064307}.

In the review, we will focus on the investigation of the nuclear magnetic moments in CDFT. In Sec. II, the CDFT framework with point-coupling interaction will be outlined briefly and the magnetic moment operator, the one-pion exchange currents and configuration mixing diagrams in first and second-order will be introduced. The magnetic moment problem in the relativistic theory is given in Sec. III. Sec. IV and Sec. V discuss the magnetic moments for nuclei with $LS$ and $jj$ closed shell plus or minus one nucleon respectively. Taking $^{33}$Mg as an example, the magnetic moments of deformed odd-$A$ nuclei will be discussed in Sec. VI. Finally, a brief summary and a perspective are given in Sec. VII.

\section{Relativistic approach for magnetic moment}

\subsection{Covariant density functional theory}

The covariant density functional theory is constructed with either the finite-range meson-exchange interaction or the contact
interaction in the point-coupling representation between nucleons~\cite{Meng2015}. For the former, the nucleus is described as
a system of Dirac nucleons that interact with each other via the exchange of mesons. For the latter, the meson exchange in each
channel (scalar¨Cisoscalar, vector¨Cisoscalar, scalar¨Cisovector, and vector¨Cisovector) is replaced by the corresponding local four-point (contact) interaction between nucleons.

Following the point-coupling representation in Ref.~\cite{Zhao2010Phys.Rev.C054319}, the basic building blocks of CDFT with point-couplings are the vertices,
\begin{equation}\label{Block}
  (\bar \psi{\cal O}\Gamma\psi),\quad {\cal O}\in\{1,\vec\tau \},\quad
  \Gamma\in\{1,\gamma_\mu,\gamma_5,\gamma_5\gamma_\mu,\sigma_{\mu\nu}\},
\end{equation}
where $\psi$ is the Dirac spinor field, $\vec{\tau}$ is the isospin Pauli matrix, and $\Gamma$ generally denotes the $4\times4$ Dirac matrices. There are 10 such building blocks characterized by their transformation properties in isospin and in Minkowski space~\cite{Zhao2010Phys.Rev.C054319}. Arrows are adopted to indicate vectors in isospin space and bold type for the
space vectors. Greek indices $\mu$ and $\nu$ run over the Minkowski indices 0, 1, 2, and 3.

A general effective Lagrangian can be written as a power
series in $\bar \psi{\cal O}\Gamma\psi$ and their derivatives~\cite{Zhao2010Phys.Rev.C054319,Burvenich2002Phys.Rev.C044308}:
\begin{eqnarray}\label{EQ:LAG}
 {\cal L} &=& \bar\psi(i\gamma_\mu\partial^\mu-m)\psi-\frac{1}{4}F^{\mu\nu}F_{\mu\nu}-e\frac{1-\tau_3}{2}\bar\psi\gamma^\mu\psi
A_\mu\nonumber\\
    & & -\frac{1}{2}\alpha_S(\bar\psi\psi)(\bar\psi\psi)
        -\frac{1}{2}\alpha_V(\bar\psi\gamma_\mu\psi)(\bar\psi\gamma^\mu\psi)\nonumber\\
    & & -\frac{1}{2}\alpha_{TV}(\bar\psi\vec{\tau}\gamma_\mu\psi)(\bar\psi\vec{\tau}
        \gamma^\mu\psi)\nonumber\\
    & & -\frac{1}{3}\beta_S(\bar\psi\psi)^3-\frac{1}{4}\gamma_S(\bar\psi\psi)^4-\frac{1}{4}
    \gamma_V[(\bar\psi\gamma_\mu\psi)(\bar\psi\gamma^\mu\psi)]^2\nonumber\\
    & & -\frac{1}{2}\delta_S\partial_\nu(\bar\psi\psi)\partial^\nu(\bar\psi\psi)
        -\frac{1}{2}\delta_V\partial_\nu(\bar\psi\gamma_\mu\psi)\partial^\nu(\bar\psi\gamma^\mu\psi) \nonumber \\
    & & -\frac{1}{2}\delta_{TV}\partial_\nu(\bar\psi\vec\tau\gamma_\mu\psi)
       \partial^\nu(\bar\psi\vec\tau\gamma_\mu\psi).
\end{eqnarray}
There are coupling constants, $\alpha_S$, $\alpha_V$,
$\alpha_{TV}$, $\beta_S$, $\gamma_S$, $\gamma_V$, $\delta_S$,
$\delta_V$, and $\delta_{TV}$. The subscripts $S$, $V$, and $T$
respectively indicate the symmetries of the couplings, i.e., scalar, vector, and isovector.

With the mean field approximation and the no-sea approximation, the nuclear energy density functional is expressed as,
\begin{equation}\label{Eq:Energy-PC}
E_{\rm DF}[\hat{\rho}]
= \int d^3\bm{r}~\mathcal{E}(\bm{r}),
\end{equation}
with the energy density
\begin{equation}\label{EDF}
 \mathcal{E}
  = \mathcal{E}_{\rm kin}(\bm{r})
    +  \mathcal{E}_{\rm int}(\bm{r})
    +  \mathcal{E}_{\rm em}(\bm{r}),
\end{equation}
which is composed of a kinetic part
\begin{equation}
   \mathcal{E}_{\rm kin}(\bm{r})
   =\sum\limits_{k=1}^A\psi^\dagger_k(\bm{r})(\balp\cdot\bp+\beta m_N-m_N)
   \psi_k(\bm{r}),
\end{equation}
where the sum over $k$ runs over the occupied orbits in the Fermi sea (no-sea approximation); an interaction part
\begin{eqnarray}
 \label{E12}
 \mathcal{E}_{\rm int}(\bm{r})
 &=& \frac{\alpha_S}{2}\rho_S^2+\frac{\beta_S}{3}\rho_S^3
    + \frac{\gamma_S}{4}\rho_S^4+\frac{\delta_S}{2}\rho_S\triangle \rho_S \nonumber\\
 &&
    + \frac{\alpha_V}{2}j_\mu j^\mu + \frac{\gamma_V}{4}(j_\mu j^\mu)^2 +
       \frac{\delta_V}{2}j_\mu\triangle j^\mu  \nonumber\\
 && +  \frac{\alpha_{TV}}{2}\vec j^{\mu}_{TV}\cdot(\vec j_{TV})_\mu+\frac{\delta_{TV}}{2}
    \vec j^\mu_{TV}\cdot\triangle(\vec j_{TV})_{\mu}, \nonumber\\
\end{eqnarray}
with the local densities and currents
  \begin{subequations}\label{Eq:dencur-PC}
  \begin{eqnarray}
  \rho_S(\bm{r})&=&\sum_{k=1}^A\bar\psi_k(\bm{r})\psi_k(\bm{r}),\\
  j^\mu(\bm{r})&=&\sum_{k=1}^A\bar\psi_k(\bm{r})\gamma^\mu\psi_k(\bm{r}),\\%
  \vec j^{\mu}_{TV}(\bm{r})
              &=&\sum_{k=1}^A\bar\psi_k(\bm{r})\gamma^\mu\vec\tau
               \psi_k(\bm{r}),
 \end{eqnarray}
 \end{subequations}
and an electromagnetic part
\begin{equation}
   \mathcal{E}_{\rm em}(\bm{r})
   = \frac{1}{4}F_{\mu\nu}F^{\mu\nu}-F^{0\mu}\partial_0A_\mu+e A_\mu j^\mu_p.
\end{equation}

Minimizing the energy density functional Eq.~(\ref{Eq:Energy-PC}) with respect to $\bar{\psi}_k$, the Dirac equation for the single nucleons is obtained,
\begin{equation}\label{Eq:Dirac-PC}
  [-i\balp\cdot\bnab+\beta\gamma_\mu V^\mu+\beta(m+S)]\psi_k(\bm{r})=\varepsilon_k\psi_k(\bm{r}).
\end{equation}
The single-particle effective Hamiltonian contains local scalar
$S(\bm{r})$ and vector $V^\mu(\bm{r})$ potentials,
\begin{equation}\label{Eq:potential-PC}
S(\bm{r})    =\Sigma_S, \quad
V^\mu(\bm{r})=\Sigma^\mu+\vec\tau\cdot\vec\Sigma^\mu_{TV},
\end{equation}
where the self-energies are given in terms of various densities,
  \begin{subequations}
  \begin{eqnarray}
  \Sigma_S           &=&\alpha_S\rho_S+\beta_S\rho^2_S+\gamma_S\rho^3_S+\delta_S\triangle\rho_S,\\
  \Sigma^\mu         &=&\alpha_Vj^\mu_V +\gamma_V (j^\mu_V)^3
                       +\delta_V\triangle j^\mu_V + e A^\mu,\\
  \vec\Sigma^\mu_{TV}&=& \alpha_{TV}\vec j^\mu_{TV}+\delta_{TV}\triangle\vec j^\mu_{TV}.
 \end{eqnarray}
 \end{subequations}

For the ground state of an even-even nucleus one has
time-reversal symmetry and the space-like parts of the
currents ${\bm{j(r)}}$ in Eq.~(\ref{Eq:dencur-PC}) as well as the vector
potential or the time-odd fields $\bm{V(r)}$ in Eq.~(\ref{Eq:potential-PC}) vanish. However, in
odd-$A$ nuclei, the odd nucleon breaks the time-reversal
symmetry, and time-odd fields give rise to a nuclear magnetic potential, which is very
important for the description of magnetic moments~\cite{Hofmann1988Phys.Lett.B307,Furnstahl1989Phys.Rev.C1398}.

Because of charge conservation in nuclei, only the third component of
isovector potential $\vec{\Sigma}^\mu_{TV}$ contributes.
The coulomb field $A_0({\bm r})$
is determined by Poisson's equation and the magnetic part ${\bm{A(r)}}$ of the electromagnetic potential is neglected in the calculation.

The relativistic residual interaction is given by the second derivative of the energy density
functional $E(\hat{\rho})$ with respect to the density matrix
\begin{equation}\label{eq:res.int.}
    V_{\alpha\beta\alpha'\beta'}
    = \frac{\delta^2E(\hat{\rho})}{\delta\hat{\rho}_{\alpha\beta}\delta\hat{\rho}_{\alpha'\beta'}}.
\end{equation}
More details can be found in Refs.~\cite{Niksic2005Phys.Rev.C014312,Daoutidis2009Phys.Rev.C024309,Li2013Phys.Rev.C064307}.

Although, because of the parity conservation, the pion meson does not contribute to the ground state in the mean field approximation at Hartree level, it plays an important role in spin-isospin excitations and is usually included in relativistic RPA and quasi-RPA calculations of these modes~\cite{DeConti1998Phys.Lett.B14,Paar2004Phys.Rev.C054303,Liang2008Phys.Rev.Lett.122502}. The widely used pion-nucleon vertex reads, in its pseudovector coupling form,
\begin{equation}\label{Eq:lag-PN}
     \mathcal {L}_{\pi N}
    = -\frac{f_\pi}{m_\pi}\bar{\psi}\gamma^\mu\gamma_5\vec{\tau}\psi\cdot\partial_\mu\vec{\pi},
\end{equation}
where $\vec{\pi}({\bm r})$ is the pion field, $f_\pi$ is the pion-nucleon coupling constant and $m_\pi$ the pion mass.

\subsection{Magnetic moment operator}

The effective electromagnetic current operator used to describe the
nuclear magnetic moment is~\cite{Furnstahl1987Nucl.Phys.A539,Furnstahl1989Phys.Rev.C1398,
Yao2006Phys.Rev.C024307,
Li2009Sci.ChinaSer.G1586,Morse1990Phys.Lett.B241},
\begin{equation}\label{electromagnetic-current}
 \hat{J}^\mu(x) =
                   Q\bar{\psi}(x)\gamma^\mu\psi(x)+\frac{\kappa}{2M}\partial_\nu
                   [\bar{\psi}(x)\sigma^{\mu\nu}\psi(x)],
\end{equation}
where the nucleon charge $Q\equiv\dfrac{e}{2}(1-\tau_3)$,
the antisymmetric tensor operator
$\sigma^{\mu\nu}=\dfrac{i} {2} [\gamma^\mu,\gamma^\nu]$, and
$\kappa$ the free anomalous gyromagnetic ratio of the nucleon with
$\kappa_p=1.793$ and $\kappa_n=-1.913$.

In Eq.~(\ref{electromagnetic-current}), the first term gives the Dirac current and second term is the so-called anomalous current. The nuclear dipole magnetic moment in units of the nuclear magneton $\mu_N=e\hbar/2Mc$,
is given by~\cite{Li2013Phys.Rev.C064307}
\begin{subequations}
\begin{eqnarray}\label{magnetic-moment}
  \mbox{\boldmath{$\mu$}}
            &=& \frac{1}{2\mu_N}\int d^3r
                \br\times\langle g.s. \vert \hat\bj(\br)\vert g.s. \rangle     \\
            &=& \int d\br\,[\frac{Mc^2}{\hbar c}Q\psi^+(\br)
                \br\times\balp\psi(\br) + \kappa\psi^+(\br)\beta\bSig\psi(\br)],  \nonumber  \\
\end{eqnarray}
\end{subequations}
where $\hat\bj(\br)$ is the operator of space-like components of the effective electromagnetic current in Eq.~(\ref{electromagnetic-current}).
The first term in above equation gives the Dirac magnetic moment, and the second term gives the anomalous magnetic moment.

Therefore, the nuclear magnetic moment operator in the relativistic theory, in units of the nuclear magneton, is
given by
\begin{eqnarray}
 \label{mm-operator}
  \mbox{\boldmath{$\hat{\mu}$}}
             &=& \frac{Mc^2}
             {\hbar c}Q\bm {r}\times\bm{\alpha}
             +\kappa\beta\bm{\Sigma}.
\end{eqnarray}

\subsection{One-pion exchange current}

\begin{figure}[h!]
\centerline{
\includegraphics[width=7.5cm]{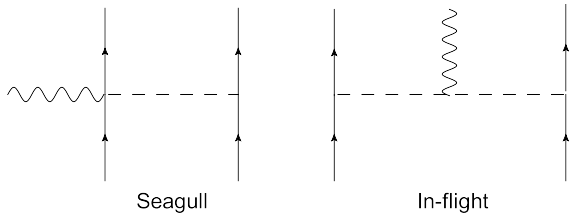}}
\caption{ Diagrams of the one-pion
exchange current: seagull (left) and in-flight (right).}
\label{jj_fig1}
\end{figure}

Although there is no explicit pion meson in CDFT at Hartree level, it is possible to study the meson exchange current corrections due to the virtual pion exchange between two
nucleons. According to Ref.~\cite{Morse1990Phys.Lett.B241}, the meson exchange current corrections are given by the two Feynman diagrams in Fig.~\ref{jj_fig1}.

As given in Ref.~\cite{Li2011Prog.Theor.Phys.1185}, the one-pion exchange current contributions to magnetic moments read,
\begin{eqnarray}\label{magnetic moment-MEC}
    \mbox{\boldmath{$\mu$}}_{\mathrm{MEC}}
    &=& \frac{1}{2}\int d \br\,\br\times
    \langle g.s.|\hat\bj^{\mathrm{seagull}}(\br)
    +\hat\bj^{\mathrm{in\mbox{-}flight}}(\br)|g.s.\rangle,\nonumber\\
\end{eqnarray}
with the corresponding one-pion exchange currents
$\hat\bj^{\mathrm{seagull}}(\br)$ and
$\hat\bj^{\mathrm{in\mbox{-}flight}}(\br)$,
\begin{widetext}
\begin{subequations}
\begin{eqnarray}
    \hat\bj^{\mathrm{seagull}}(\br)
    &=&-\frac{8ef^2_{\pi}M}{m^2_\pi} \int d \bm {x}\,
        \bar{\psi}_p(\svec r) {\bm\gamma}\gamma_5\psi_n(\svec r)
         D_\pi(\svec r,\svec x)
    \bar{\psi}_n(\svec x)\frac{M^*}{M}\gamma_5\psi_p(\svec x),\\
    \hat\bj^{\mathrm{in\mbox{-}flight}}(\br)
    &=&-\frac{16ief^2_{\pi}M^2}{m_\pi^2} \int d\svec x d\svec y \bar{\psi}_p(\svec x)\frac{M^*}{M}\gamma_5\psi_n(\svec x)
        D_\pi(\svec x,\svec r){\bm\nabla}_{\svec r}
        D_\pi(\svec r,\svec y)\times\bar{\psi}_n(\svec y)\frac{M^*}{M}\gamma_5\psi_p(\svec
        y).
\end{eqnarray}
\end{subequations}
\end{widetext}

The pion propagator in $r$ space has the form,
\begin{equation}
  D_\pi(\svec x,\svec r)=\dfrac{1}{4\pi}\dfrac{e^{-m_\pi|\svec
x-\svec r|}} {|\svec x-\svec r|}.
\end{equation}

\subsection{First-order corrections}

The residual interaction, neglected in the mean field approximation, leads to configuration mixing, i.e., the coupling between the valence nucleon and particle-hole states in the core. It is also called core polarization. The configuration mixing corrections to the magnetic moment are treated approximately by the perturbation theory.

\begin{figure}
\centerline{
\includegraphics[width=8.5cm]{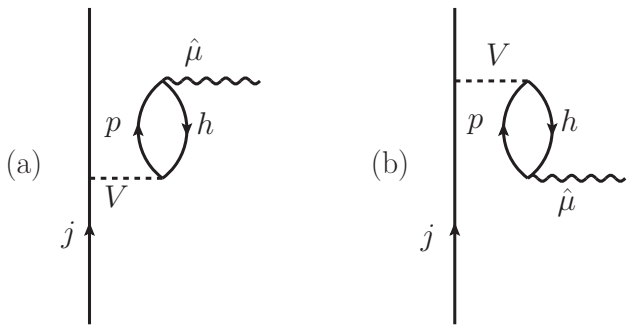}}
\caption{ Diagrams of first-order configuration mixing corrections to the magnetic moment.
The external line represents the valence nucleon, and the intermediate
particle-hole pair represents an excited state of the core.}
\label{fig2}
\end{figure}

According to the perturbation theory, the
first-order correction to the magnetic moments is given by
\begin{eqnarray}\label{eq:1st}
  \delta\mu_{\mathrm{1st}}
  &=& \langle n|\hat{\mu}\frac{\hat{Q}}{E_n-\hat{H}_0}\hat{V}|n\rangle
  + \langle n|\hat{V}\frac{\hat{Q}}{E_n-\hat{H}_0}\hat{\mu}|n\rangle,\nonumber\\
\end{eqnarray}
where $\vert n\rangle$ and $E_n$ denote the ground-state wavefunctions and corresponding energies, $\hat H_0$ and $\hat V$ are respectively the operators of the mean-field Hamiltonian and the residual interaction, $\hat{\mu}$ is the magnetic moment operator, and $\hat Q$ projects onto multi-particle and multi-hole configurations.

The corresponding Feynman diagram is shown in Fig.~\ref{fig2}, where the wiggly lines represent the external field (here the magnetic moment operator), and the dashed lines denote the residual interaction. Solid lines with arrow upwards denote particle states (i.e., single particle orbits above the Fermi surface) and those with arrow downwards are hole states (i.e., single particle orbits in the Fermi sea).

For the magnetic moments of the nuclei with a doubly closed shell core $\pm1$ nucleon, the first-order correction can be simplified as~\cite{Li2013Phys.Rev.C064307},
\begin{widetext}
\begin{equation}\label{eq:1cp-bq}
    \delta\mu_{1\mathrm{st}}=\sum_{j_pj_hJ}
    \frac{2\langle j_h\|\bmu\|j_p\rangle}{\Delta E_j}
    (-1)^{j_h+j+J}\hat{j\,}^{-1}\sqrt{\frac{j}{j+1}}(2J+1)
    \left\{
      \begin{array}{ccc}
        j_h & j_p & 1 \\
        j & j & J \\
      \end{array}
    \right\}
    \langle jj_p;JM|V|jj_h;JM\rangle,
\end{equation}
\end{widetext}
in which $j$ denotes the valence nucleon state, $j_p$ and $j_h$ are respectively for particle and hole states,
and $\Delta E = \varepsilon_{j_p}-\varepsilon_{j_h}$ is the excitation energy of the one-particle-one-hole (1p-1h) excitation.

The selection rule $\Delta\ell=0$ of the non-relativistic magnetic moment operator allows only particles and holes as spin-orbit
partners, i.e., $j_p=\ell-\frac{1}{2}$ and $j_h=\ell+\frac{1}{2}$. All other diagrams vanish.
Therefore the first-order configuration mixing
does not provide any contribution in nuclei with an $LS$ closed core $\pm1$ nucleon, because there are no spin-orbit
partners on both sides of the Fermi surface and the magnetic moment operator cannot couple to
magnetic resonances~\cite{Arima1987Adv.Nucl.Phys.1}.

\subsection{Second-order corrections}

\begin{figure}[h]
\centerline{
\includegraphics[width=8.5cm]{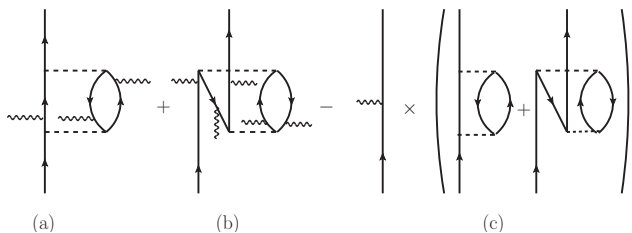}}
\caption{ Diagrams of second-order configuration mixing corrections to magnetic moment:
(a) 1p-1h mode, (b) 2p-2h mode and (c) wave function renormalization. Diagrams with more than one external wiggly
lines are an abbreviation for several separate diagrams where each of them has only one wiggly line at the indicated places.}
\label{jj_fig3}
\end{figure}

As shown in Refs.~\cite{Shimizu1974Nucl.Phys.A282,Arima1987Adv.Nucl.Phys.1,Li2011Sci.ChinaPhys.Mech.Astron.204}, the second-order correction to the magnetic moments is given by
\begin{eqnarray}\label{eq:sec}
  \delta\mu_{\mathrm{2nd}}
   &=& \langle n|\hat V\frac{\hat Q}{E_n-\hat H_0}\hat{\mu}\frac{\hat Q}{E_n-\hat H_0}\hat V|n\rangle\nonumber\\
   & & -\langle n|\hat{\mu}|n\rangle
        \langle
        n|\hat V\frac{\hat Q}{(E_n-\hat H_0)^2}\hat V|n\rangle,
\end{eqnarray}
where the second term comes from the renormalization of nuclear wave
function.

The second-order corrections include one-particle-one-hole (1p-1h) and two-particle-two-hole (2p-2h) contributions.
As shown in Fig.~\ref{jj_fig3}, the second-order correction to the magnetic moment for a nucleus with a doubly closed shell core plus one nucleon can be divided into three terms~\cite{Arima1987Adv.Nucl.Phys.1}, the contributions of two-particle-one-hole (2p-1h) configurations, three-particle-two-hole (3p-2h) configurations,
and the wave function renormalization respectively.

\begin{figure}
 \centering
 \includegraphics[width=8.5cm]{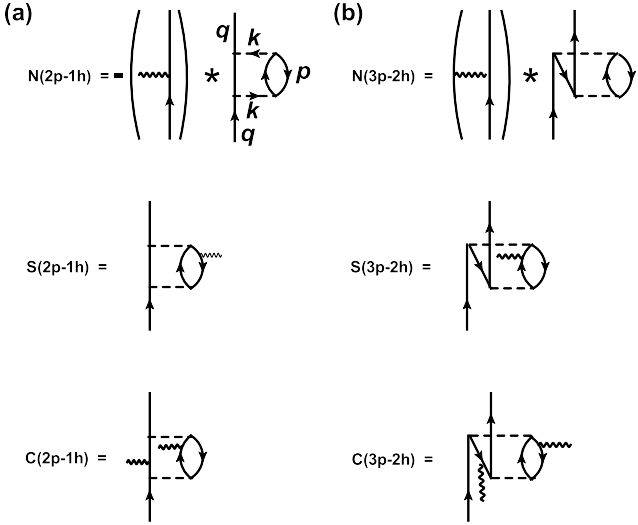}\\
 \caption{Diagrams representing second-order configuration mixing
corrections: (a) 2p-1h and (b) 3p-2h intermediate states.
For the notations N, S, and C, see the text for details.
\label{jj_fig4}}
\end{figure}

As shown in Fig.~\ref{jj_fig4}, for a nucleus with a doubly closed core plus one particle, the corresponding second-order corrections to the magnetic moment include,
\begin{itemize}
  \item  N(2p-1h) and N(3p-2h): from the wave function renormalization.
  \item  S(2p-1h) and C(3p-2h): the external field operator acting on the hole line.
  \item  C(2p-1h) and S(3p-2h): the external field operator acting on the particle
  line,
\end{itemize}
and the second-order correction in this case,
\begin{eqnarray}
    \delta\mu_{\mathrm{2nd}}
   &=& ~\mbox{N(2p-1h)} + \mbox{S(2p-1h)} + \mbox{C(2p-1h)}\nonumber\\
   & & + \mbox{N(3p-2h)} + \mbox{S(3p-2h)} + \mbox{C(3p-2h)}.
\end{eqnarray}

Accordingly, for a nucleus with a doubly closed core minus one nucleon, it reads
\begin{eqnarray}
    \delta\mu_{\mathrm{cm}}^{\mathrm{2nd}}
   &=& ~\mbox{N(2h-1p)} + \mbox{S(2h-1p)} + \mbox{C(2h-1p)}\nonumber\\
   & & + \mbox{N(3h-2p)} + \mbox{S(3h-2p)} + \mbox{C(3h-2p)}.
\end{eqnarray}

\section{Magnetic moment problem in relativistic theory}

Although CDFT has been successfully applied to investigate the nuclear structure over
the whole periodic table, from light to superheavy nuclei with a few
universal parameters~\cite{Ring1996Prog.Part.Nucl.Phys.193,
Vretenar2005Phys.Rep.101,
Meng2006Prog.Part.Nucl.Phys.470,
Meng2011Prog.Phys.199,Niksic2011Prog.Part.Nucl.Phys.519,
Meng2015}, a long-standing problem is its failure to predict magnetic moments.

It is already known from the very beginning that the straightforward applications of the
relativistic single-particle model, where only sigma and the time-like component of the vector mesons were considered, will lead to a  significantly larger isoscalar magnetic moments than the observed
values~\cite{Miller1975Ann.Phys.40,Serot1981Phys.Lett.B263}. This is because the reduced Dirac effective mass of the nucleon enhances the relativistic effect on the electromagnetic current~\cite{McNeil1986Phys.Rev.C746}.
It can be easily understood as follows.

From the Gordon identity, the Dirac current from the Dirac equation with scalar and vector potentials can be decomposed into an orbital (conventional) current and a spin current,
\begin{eqnarray}\label{current-Dirac}
 \bm{j}_D
 &=& Q\bar{\psi}(\bm{r})\bm{\gamma}\psi(\bm{r}) \nonumber  \\
 &=& \frac{Q}{M^*}\bar{\psi}(\bm{r})\bm{p}\psi(\bm{r})
    + \frac{Q}{2M^*}\bm{\nabla}\times[\psi^+(\bm{r})\beta\bm{\Sigma}\psi(\bm{r})], \nonumber\\
\end{eqnarray}
with the nucleon effective (scalar) mass $M^*=M+S\,\approx0.6M$.

Therefore the Dirac magnetic moment,
\begin{eqnarray}
  \bm{\mu}_D &=& \frac{1}{2\mu_N}\int d\bm{r}\,
                \br\times \bm{j}_D(\bm{r})\nonumber\\
             &=& Q\int d\bm{r}\frac{M}{M^*}\bar{\psi}[\bm{L}+\bm{\Sigma}]\psi,
\end{eqnarray}
is enhanced.

It was pointed out that the valence-nucleon approximation is wrong in the relativistic calculation~\cite{Serot1992Rep.Prog.Phys.55.1855}.
In reality, the current entering the magnetic moment operator is not just the single particle current of the single valence nucleon. The valence nucleon will polarize the surrounding medium and lead to a polarization current induced by the isoscalar current-current interaction. The effective current is therefore reduced. This so-called ``back-flow" effect has been treated in the literature in different ways.

The induced current has been calculated in infinite nuclear matter by a Ward identity~\cite{Bentz1985Nucl.Phys.A593} or in the framework of Landau-Migdal theory~\cite{McNeil1986Phys.Rev.C746} and the results have been applied in a local density approximation to finite nuclei. This has been improved by calculating the polarization current in linear response theory in finite spherical systems summing up the loop-diagrams~\cite{Ichii1987Phys.Lett.B11,Shepard1988Phys.Rev.C1130,Furnstahl1988PhysRevC.38.370}.

The most direct method, however, is to treat the finite odd-$A$ system in a fully self-consistent way~\cite{Hofmann1988Phys.Lett.B307,Furnstahl1989Phys.Rev.C1398,Yao2006Phys.Rev.C024307,Li2009Sci.ChinaSer.G1586}.
The valence nucleon sits in a certain sub-shell with the magnetic quantum number $m$. This leads to a small axially symmetric deformation and an azimuthal current $j_\varphi$ around the symmetry axis, which induces in the core a nuclear magnetic field, i.e., the time-odd fields, and the corresponding polarization currents. This can be taken into account in a fully self-consistent way by solving the deformed Dirac equations in CDFT with nuclear magnetic fields breaking time-reversal symmetry. For odd-$A$ nuclei in the direct vicinity of an $LS$ closed core, all these methods lead to a strong reduction of the effective current such that the enhancement due to the small effective Dirac mass is nearly canceled. Finally, the resulting isoscalar magnetic moments are in excellent with the Schmidt values of the conventional non-relativistic single particle model.

\begin{figure}
\centerline{
\includegraphics[width=7.5cm]{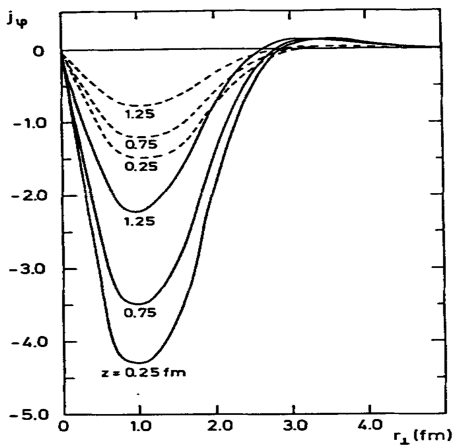}}
\caption{ The azimuthal current $j_\varphi(z, r_\bot)$ in the nucleus $^{15}$N as a function of the radial distance $r_\bot$ for different values of the coordinate $z$ along the symmetry axis. Dashed lines are calculated without polarization of the $^{16}$O core, full lines include in addition the polarization of the core. The units are $10^{-3}$ fm$^{-3}$. Reproduced from Ref.~\cite{Hofmann1988Phys.Lett.B307}.}
\label{current_N15}
\end{figure}

In Fig.~\ref{current_N15}, the azimuthal current $j_\varphi(z, r_\bot)$ for the nucleus $^{15}$N is shown after self-consistently considering the polarization current in deformed CDFT with time-odd fields~\cite{Hofmann1988Phys.Lett.B307}. It is purely azimuthal. The dashed lines are obtained without polarization, i.e.,
removing from the self-consistently determined $^{16}$O core only one proton in the $1p_{1/2}$ shell. While the full lines are obtained from a self-consistent solution for the odd system including
the spatial part of the $\omega$ field (Nuclear Magnetism). The external hole induced a current in the core,
which enhances the total current by nearly a factor three.

\begin{table}\tabcolsep=10pt
  \centering
  \caption{The isoscalar magnetic moments for the particle numbers $A$ =
15, 17, 39 and 41. Covariant density functional theory calculations
for a spherical core, i.e., without polarization currents. Reproduced from
Ref.~\cite{Hofmann1988Phys.Lett.B307}.}\label{Isoscalar-def}
  \begin{tabular}{ccccc}
     \hline\hline
     A & 15 & 17 & 39 & 41 \\ \hline
     \multicolumn{5}{c}{magnetic moments $\mu$}\\ \hline
     Exp. & 0.22 & 1.41 & 0.71 & 1.92 \\
     Schmidt & 0.19 & 1.44 & 0.64 & 1.94 \\
     Sph.CDFT & 0.32 & 1.57 & 0.94 & 2.21 \\
     Def.CDFT & 0.18 & 1.48 & 0.64 & 1.97 \\ \hline
     \multicolumn{5}{c}{Dirac part $\mu_D$}\\ \hline
     Schmidt & 0.17 & 1.50 & 0.60 & 2.00 \\
     Sph.CDFT & 0.30 & 1.63 & 0.91 & 2.27 \\
     Def.CDFT & 0.15 & 1.54 & 0.60 & 2.03 \\ \hline
     \multicolumn{5}{c}{Anomal part $\mu_A$}\\ \hline
     Schmidt & 0.02 & -0.06 & 0.04 & -0.06 \\
     Sph.CDFT & 0.02 & -0.06 & 0.04 & -0.06 \\
     Def.CDFT & 0.03 & -0.06 & 0.04 & -0.06 \\
     \hline
   \end{tabular}
\end{table}

Table~\ref{Isoscalar-def} shows isoscalar magnetic moments for the odd mass nuclei with the mass numbers $A$ = 15,
17, 39, and 41.
Experimental values are compared with the non-relativistic Schmidt values, with those of the spherical CDFT where only the doubly magic core is determined self-consistently (Sph.CDFT), and values and with fully self-consistent calculations
in the odd-mass system including time odd fields (Def.CDFT).
It is observed that the anomalous part of the isoscalar magnetic moments is extremely small in all cases. Because of the negative contributions of the polarization currents in the core to the Dirac moment for both neutron nuclei and proton nuclei, excellent agreement is found between the Schmidt values and the deformed CDFT results.

It can therefore be concluded, that it is not the mean field approximation which produces such poor results for the isoscalar magnetic moments in relativistic theories. If the time reversal symmetry violation in the wave function of the odd mass nuclei and the corresponding polarization currents are taken into account fully self-consistently, the relativistic theory is able to describe the Schmidt values properly.

\section{Magnetic moments for nuclei with $LS$ closed shell core $\pm1$ nucleon}

The magnetic moments for nuclei with $LS$ closed shell core $\pm1$ nucleon are of particular importance, because there are no spin-orbit partners on both sides of the fermi surface and therefore all first-order configuration mixing corrections vanish. In Refs.~\cite{Li2011Sci.ChinaPhys.Mech.Astron.204,Li2011Prog.Theor.Phys.1185, Wei2012Prog.Theor.Phys.Supp.400}, the contributions from one-pion exchange current (MEC) and then the second-order corrections (2nd) are presented. Here, the isoscalar and isovector magnetic moments of light odd-mass nuclei near the $LS$ closed shells with $A = 15$, 17, 39 and 41 will be discussed.

\begin{table}[h!]\tabcolsep=3pt
\caption{The one-pion exchange current corrections to
the isovector magnetic moments obtained from CDFT calculations using
PK1 and PC-F1 effective interaction, in comparison with the QHD-II~\cite{Morse1990Phys.Lett.B241} and non-relativistic
results~\cite{Chemtob1969Nucl.Phys.A449,Hyuga1980Nucl.Phys.A363,Towner1983Nucl.Phys.A334,Ito1987Ann.Phys.169}. Reproduced from Refs.~\cite{Li2011Prog.Theor.Phys.1185,Li2011Sci.ChinaPhys.Mech.Astron.204}. }
\label{tab:mec}
\begin{tabular}{ccccccccc}
  \hline\hline
\multirow{2}{*}{~~A~~} &\multicolumn{4}{c}{Non-relativistic}&
&\multicolumn{3}{c}{Relativistic}
\\ \cline{2-5}\cline{7-9}
  &\cite{Chemtob1969Nucl.Phys.A449} & \cite{Hyuga1980Nucl.Phys.A363}& \cite{Towner1983Nucl.Phys.A334}&     \cite{Ito1987Ann.Phys.169} && \cite{Morse1990Phys.Lett.B241} & PK1\cite{Li2011Prog.Theor.Phys.1185} & PC-F1\cite{Li2011Sci.ChinaPhys.Mech.Astron.204}
\\ \hline
15 & 0.127 & 0.116& 0.092&  0.111 && 0.102 & 0.091 & 0.091 \\
17 & 0.084 & 0.093& 0.065&  0.092 && 0.151 & 0.092 & 0.093  \\
39 & 0.204 & 0.199& 0.149&  0.184 && 0.174 & 0.190 & 0.190  \\
41 & 0.195 & 0.201& 0.115&  0.180 && 0.270 & 0.184 & 0.184  \\
  \hline
\end{tabular}
\end{table}

The one-pion exchange current corrections to the isovector magnetic
moments obtained from CDFT calculations using PK1~\cite{Long2004Phys.Rev.C034319} and PC-F1~\cite{Burvenich2002Phys.Rev.C044308} effective interactions are compared in Table~\ref{tab:mec} with QHD-II
claculations~\cite{Morse1990Phys.Lett.B241} and non-relativistic
calculations~\cite{Chemtob1969Nucl.Phys.A449,Hyuga1980Nucl.Phys.A363,Towner1983Nucl.Phys.A334,Ito1987Ann.Phys.169}.

It is pointed out early~\cite{Chemtob1969Nucl.Phys.A449,Yamazaki1970PhysRevLett.25.547} that the one-pion exchange currents give mainly an enhancement (reduction) of the orbital $g$ factor for proton (neutron), which always give a positive contribution to the isovector magnetic moments. This is consistent with the relativistic calculations. It is shown that the obtained corrections to the isovector magnetic moments in relativistic theories are in reasonable agreement with other
calculations. As noted in Ref.~\cite{Morse1990Phys.Lett.B241}, the differences
between the various calculations are most likely due to relative small changes in the balance of
contributions from seagull and in-flight diagrams rather than any
fundamental differences in the models used. It is also confirmed by other relativistic
effective interaction, and almost the same results are obtained.

\begin{table*}\tabcolsep=5pt
\caption{\label{tab:isoscalar}Isoscalar magnetic moments obtained
from CDFT calculations using PC-F1~\cite{Burvenich2002Phys.Rev.C044308} effective interaction, in comparison
with the corresponding data, Schmidt value (Sch.), previous relativistic
results~\cite{Morse1990Phys.Lett.B241} and non-relativistic
results~\cite{Towner1987Phys.Rep.263,Arima1987Adv.Nucl.Phys.1}. Reproduced from Refs.~\cite{Li2011Prog.Theor.Phys.1185,Li2011Sci.ChinaPhys.Mech.Astron.204}. }
\begin{tabular}{cccccccccccc}
  \hline\hline
\multirow{2}{*}{A} & \multirow{2}{*}{Exp.}
&\multicolumn{6}{c}{Non-relativistic}&
&\multicolumn{3}{c}{Relativistic}
\\ \cline{3-8}\cline{10-11}
 & &Sch. &\multicolumn{2}{c}{Sch.+MEC}&&\multicolumn{2}{c}{Sch.+MEC+2nd} && QHD+MEC && CDFT+MEC
\\  \cline{4-5}\cline{7-8}\cline{10-10}\cline{12-12}
&&&~\cite{Towner1987Phys.Rep.263}~&~\cite{Arima1987Adv.Nucl.Phys.1}~&&~\cite{Towner1987Phys.Rep.263}~&~\cite{Arima1987Adv.Nucl.Phys.1}~&&~\cite{Morse1990Phys.Lett.B241} & &PC-F1\\ \hline
15  &~0.218~ &~0.187~ &~0.194~&~0.190~ &&~0.228~ &0.233 & & 0.200(0.199$+$0.001)&& ~$0.216(0.216+0.000)$~\\
17  &~1.414~ &~1.440~ &~1.439~&~1.478~ &&~1.410~ &1.435 & & 1.42~(1.43~$-$0.011)&& ~$1.465(1.467-0.002)$~\\
39  &~0.706~ &~0.636~ &~0.640~&~0.655~ &&~0.706~ &0.735 & & 0.659(0.660$-$0.001)&& ~$0.707(0.707+0.000)$~\\
41  &~1.918~ &~1.940~ &~1.936~&~1.996~ &&~1.893~ &1.944 & & 1.93~(1.94~$-$0.007)&& ~$1.978(1.985-0.007)$~\\
  \hline
\end{tabular}
\end{table*}

Table~\ref{tab:isoscalar} presents the isoscalar magnetic moments and
corresponding pion exchange current corrections obtained from non-relativistic
results~\cite{Towner1987Phys.Rep.263,Arima1987Adv.Nucl.Phys.1}, previous relativistic
result~\cite{Morse1990Phys.Lett.B241}, and CDFT calculations using PC-F1 as well as the Schmidt values and corresponding data.
As discussed in Ref.~\cite{Li2011Prog.Theor.Phys.1185}, the
CDFT results are in good agreement with data, and
in some case better than the QHD~\cite{Morse1990Phys.Lett.B241}
and non-relativistic calculations~\cite{Towner1987Phys.Rep.263,Arima1987Adv.Nucl.Phys.1}.
Moreover, same as in Ref.~\cite{Morse1990Phys.Lett.B241}, the MEC corrections to
isoscalar moments in CDFT calculations are negligible. For the mirror
nuclei with double-closed shell plus or minus one nucleon, the MEC
corrections to isoscalar moments reflect the violation of isospin
symmetry in wave functions. With the small MEC corrections to
isoscalar moments here, it is easy to understand the excellent
description of the isoscalar magnetic moments in deformed CDFT with space-like components of vector meson in
Refs.~\cite{Hofmann1988Phys.Lett.B307,Yao2006Phys.Rev.C024307}.

\begin{table*}\tabcolsep=5pt
\caption{\label{tab:isovector} Similar as Table~\ref{tab:isoscalar},
but for the isovector magnetic moments. The isovector magnetic
moments without the pion exchange current corrections in QHD and CDFT
calculations are given in the brackets respectively. Reproduced from Refs.~\cite{Li2011Prog.Theor.Phys.1185,Li2011Sci.ChinaPhys.Mech.Astron.204}. }
\begin{tabular}{cccccccccccc}
  \hline\hline
\multirow{2}{*}{~~A~~} & \multirow{2}{*}{Exp.}
&\multicolumn{6}{c}{Non-relativistic}&
&\multicolumn{3}{c}{Relativistic}
\\ \cline{3-8}\cline{10-12}
 & &Sch. &\multicolumn{2}{c}{Sch.+MEC}&&\multicolumn{2}{c}{Sch.+MEC+2nd} && QHD+MEC & & CDFT+MEC
\\  \cline{4-5}\cline{7-8}\cline{10-10}\cline{12-12}
&&&~\cite{Towner1987Phys.Rep.263}~&~\cite{Arima1987Adv.Nucl.Phys.1}~&&~\cite{Towner1987Phys.Rep.263}~&~\cite{Arima1987Adv.Nucl.Phys.1}~&&~\cite{Morse1990Phys.Lett.B241} & &PC-F1\\ \hline
15  &$-0.501$ &$-0.451$&$-0.427$&$-0.483$  &&$-0.456$ & $-0.508$ & &$-0.347(-0.449)$& &$-0.339(-0.430)$  \\
17  &$3.308$ &$3.353$&$3.729$&$3.873$ & &$3.281$&$3.306$  & &$3.61(3.46)$ & & $3.576(3.484)$ \\
39  &$-0.316$ &$-0.512$  &$-0.149$&$-0.265$  &&$-0.286$&$-0.481$ & &$-0.106(-0.280)$ &          & $-0.115(-0.305)$  \\
41  &$3.512$ &$3.853$ &$4.621$&$4.492$  &&3.803&3.729 & &$4.41(4.14)$ & & $4.322(4.138)$ \\
  \hline
\end{tabular}
\end{table*}

In Table~\ref{tab:isovector}, the corresponding results for isovector magnetic moments are presented. As discussed in Ref.~\cite{Li2011Prog.Theor.Phys.1185}, the pion exchange current gives a significant
positive correction to isovector magnetic moments, which is
consistent with the calculations in Ref.~\cite{Morse1990Phys.Lett.B241}. Compared
with the case for the isoscalar magnetic moments, however, the
relativistic isovector magnetic moments calculated with pion
exchange current corrections deviate much more from data than the
calculation without pion exchange current corrections. This
phenomenon is also found in CDFT calculations with other effective
interactions. It can be understood as follows.

In CDFT the magnetic moments are enhanced due
to the small effective nucleon mass by the scalar field, which is
cancelled by the space-like components of vector field (vertex correction~\cite{Arima1987Adv.Nucl.Phys.1} or
the consideration of space-like component of $\omega$
meson~\cite{Hofmann1988Phys.Lett.B307}) for the isoscalar parts, but not for the
isovector parts~\cite{Arima2011Sci.ChinaPhys.Mech.Astron.188}. As shown in
Table~\ref{tab:isovector}, the CDFT values without pion are always
larger than the Schmidt values, which reflects the $1/M^*$ enhancement
of the orbital isovector $g$-factor. Then the pion gives additional
enhancements, which leads to deviate too much from the isovector
magnetic moment in the end. Therefore, the CDFT with only one-pion
exchange current correction enhances the isovector magnetic moments
further and does not improve the corresponding description for the
concerned nuclei.

From Sch.+MEC+2nd (column 6 and 7) in Table~\ref{tab:isovector}, it is shown that the second-order configuration mixing effects included in the non-relativistic calculations have canceled the enhancement effect
of MEC and the net effect of them gives the right sign for the
correction to the Schmidt isovector magnetic moments. It is expected
that the second-order corrections may improve the description of
isovector magnetic moments in CDFT.

In Ref.~\cite{Li2011Sci.ChinaPhys.Mech.Astron.204}, the second-order corrections to magnetic moments in $^{17}$O,
$^{17}$F, $^{41}$Ca, $^{41}$Sc, $^{15}$O, $^{15}$N, $^{39}$Ca, and
$^{39}$K are also calculated in CDFT with the PC-F1~\cite{Burvenich2002Phys.Rev.C044308}.
In Table~\ref{tab:2nd-LS}, the second-order corrections to the magnetic
moments in CDFT are compared with previous non-relativistic
results~\cite{Shimizu1974Nucl.Phys.A282,Mavromatis1967Nucl.Phys.A17,Horie1973,Ichimura1965}.

\begin{table}\tabcolsep=3pt
\caption{\label{tab:2nd-LS} Second-order configuration mixing corrections to the magnetic moments in CDFT with PC-F1~\cite{Burvenich2002Phys.Rev.C044308} in comparison with previous non-relativistic results~\cite{Shimizu1974Nucl.Phys.A282,Mavromatis1967Nucl.Phys.A17,Horie1973,Ichimura1965}. Reproduced from Ref.~\cite{Li2011Sci.ChinaPhys.Mech.Astron.204}.}
\begin{threeparttable}
\renewcommand{\thefootnote}{\fnsymbol{footnote}} 
\begin{tabular}{ccccccccc}
  \hline\hline
\multicolumn{2}{c}{\multirow{2}{*}{Orbit}}
&\multicolumn{5}{c}{Non-relativistic}&
&\multicolumn{1}{c}{Relativistic}\\ \cline{3-7}\cline{9-9}
  &&SIA\tnote{1} & MZ\tnote{2} & MZB\tnote{3} &     IY(I)\tnote{4} &  TY\tnote{4} &&
  \cite{Li2011Sci.ChinaPhys.Mech.Astron.204}
\\ \hline
$^{15}$N &$\pi1p^{-1}_{1/2}$ &-0.06 & -0.06&-0.10& -0.09&-0.12  &&-0.086 \\
$^{15}$O &$\nu1p^{-1}_{1/2}$ &~0.10 & ~0.14&~0.10& ~0.09&~0.12  &&-0.053  \\
$^{17}$F &$\pi1d_{5/2}$ &-0.21 & -0.30&-0.16& -0.20&-0.27       &&-0.377  \\
$^{17}$O &$\nu1d_{5/2}$ &~0.20 & ~0.28&~0.16& ~0.20&~0.27       &&~0.423  \\
$^{39}$K &$\pi1d^{-1}_{3/2}$ &-0.21 &      &-0.26&      &       &&-0.161 \\
$^{39}$Ca &$\nu1d^{-1}_{3/2}$&~0.26 &      &~0.26&      &       &&-0.099  \\
$^{41}$Sc &$\pi1f_{7/2}$&-0.41 &      &-0.28& -0.33&-0.42       &&-0.627  \\
$^{41}$Ca &$\nu1f_{7/2}$&~0.40 &      &~0.28& ~0.33&~0.42       &&~0.681  \\
  \hline
\end{tabular}
\begin{tablenotes}
 \footnotesize
 \item[1]G-matrix from the Reid potential for the $^{16}$O region; G-matrix from the
Hamada-Johnston potential for the $^{40}$Ca region~\cite{Shimizu1974Nucl.Phys.A282} .
 \item[2]G-matrix from the
Hamada-Johnston potential~\cite{Mavromatis1967Nucl.Phys.A17}.
 \item[3]The Kallio-Kolltveit potential~\cite{Horie1973} .
 \item[4]Two kinds of phenomenological potential~\cite{Ichimura1965}.
\end{tablenotes}
\end{threeparttable}
\end{table}

Table~\ref{tab:2nd-LS} shows that the relativistic second-order corrections are of the same order and have the
same sign as the non-relativistic results, with the exception of
$^{15}$O and $^{39}$Ca. According to formulas for second-order corrections, the differences between
the relativistic and the non-relativistic results are mainly ascribed
to the magnetic moment operator and to the residual
interactions.

\begin{table}
\tabcolsep=6pt
\caption{\label{tab:2nd-ph} Contributions from different excitation modes in the
relativistic calculations, in comparison with the previous
non-relativistic calculations~\cite{Shimizu1974Nucl.Phys.A282}. Reproduced from Ref.~\cite{Li2011Sci.ChinaPhys.Mech.Astron.204}.}
\begin{tabular}{ccccccc}
  \hline\hline
&&\multicolumn{2}{c}{Non-relativistic}&&\multicolumn{2}{c}{Relativistic}\\
\cline{3-4}\cline{6-7}
 Orbit & Mode &$\delta\langle\bm\mu^{(0)}\rangle$ & $\delta\langle\bm\mu^{(1)}\rangle$
       && $\delta\langle\bm\mu^{(0)}\rangle$ & $\delta\langle\bm\mu^{(1)}\rangle$
\\ \hline
\multirow{2}{*}{$1p^{-1}_{1/2}$}
 &2h-1p &  0.014& 0.075&   & -0.006&-0.022 \\
 &3h-2p &  0.003& 0.005&   &  0.063&-0.005 \\ \hline
\multirow{2}{*}{$1d_{5/2}$}
 &2p-1h & -0.004& 0.138&   &  0.003&-0.028 \\
 &3p-2h &  0    & 0.070&   & -0.020& 0.373 \\ \hline
\multirow{2}{*}{$1d^{-1}_{3/2}$}
 &2h-1p &  0.017& 0.183&   & -0.017&-0.062 \\
 &3h-2p &  0.005& 0.052&   &  0.113&-0.031 \\ \hline
\multirow{2}{*}{$1f_{7/2}$}
 &2p-1h & -0.007& 0.261&   &  0.010&-0.062  \\
 &3p-2h & -0.001& 0.140&   & -0.017& 0.592  \\
  \hline
\end{tabular}
\end{table}

For a further comparison, the contributions from different
kinds of excitation modes (2p-1h, 3p-2h, etc.) both for
relativistic and non-relativistic calculations~\cite{Shimizu1974Nucl.Phys.A282} are
tabulated in Table~\ref{tab:2nd-ph}. $\bm\mu^{(0)}$ and $\bm\mu^{(1)}$ represent the
isoscalar and isovector magnetic moments, respectively. As discussed in Ref.~\cite{Li2011Sci.ChinaPhys.Mech.Astron.204}, 2p-1h and 2h-1p excitation modes give the main contribution
in the non-relativistic results, while 3p-2h and 3h-2p
excitation modes give the main contribution to the relativistic
results.

\begin{figure}[!ht]
 \centering
\includegraphics[width=8cm]{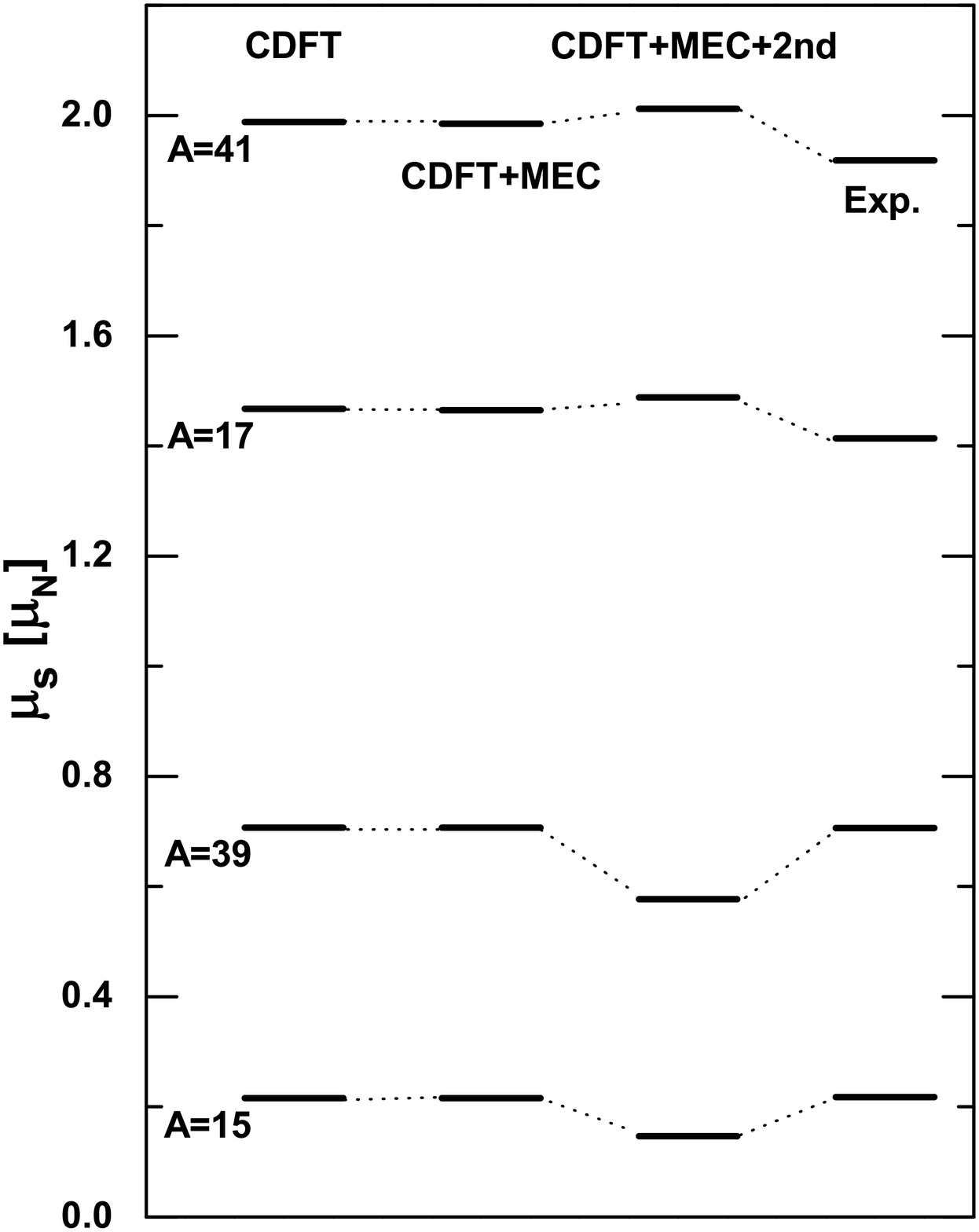}
 \vspace*{-0.9cm}\\
 \caption{\label{fig:3} Isoscalar magnetic moments obtained in CDFT calculation with MEC and present second-order corrections (2nd), in comparison with the corresponding data. Reproduced from Ref.~\cite{Li2011Sci.ChinaPhys.Mech.Astron.204}.}
\end{figure}

\begin{figure}[!ht]
 \centering
 \includegraphics[width=8cm]{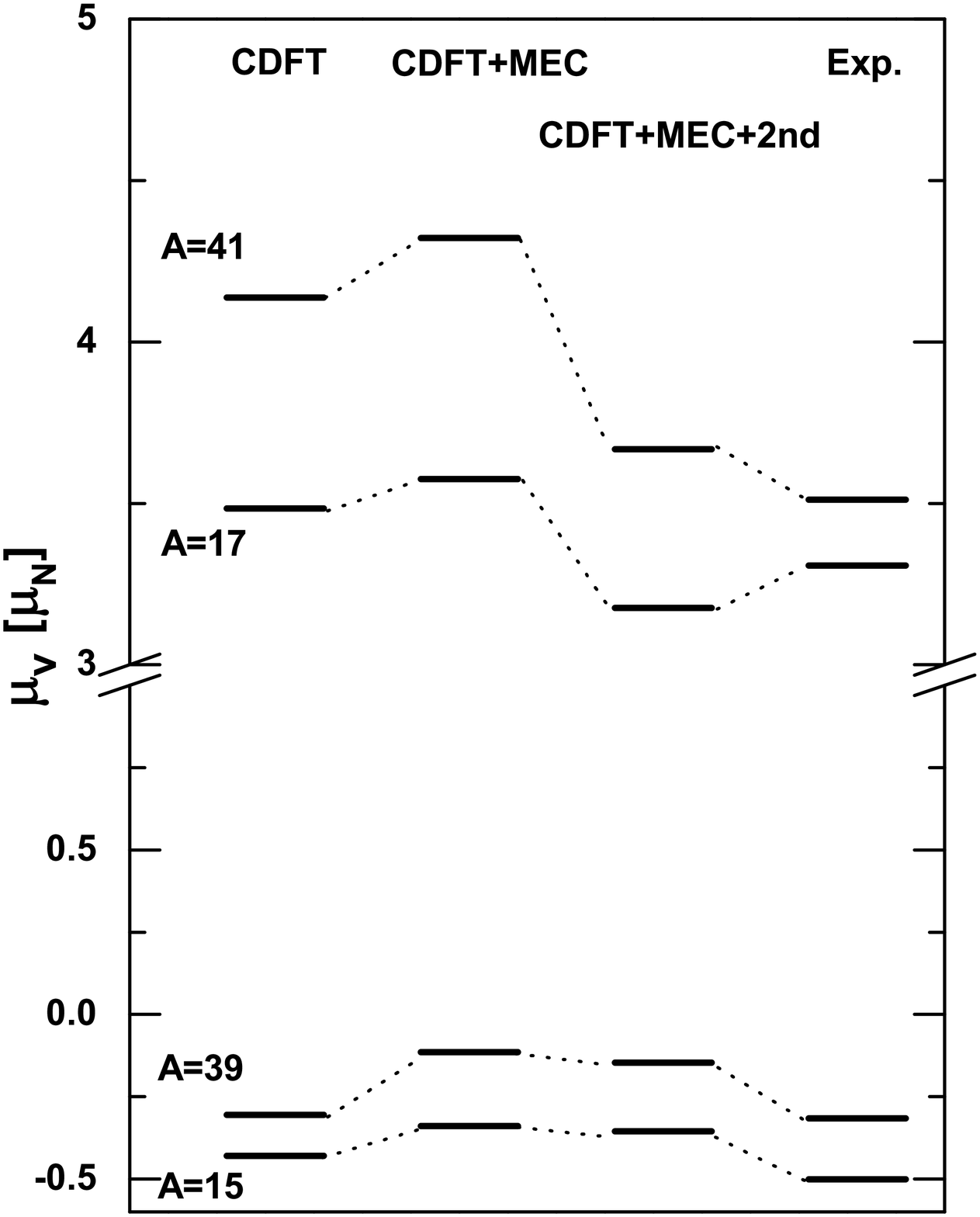}\vspace*{-0.9cm}\\
 \caption{\label{fig:4} Same as Fig.~\ref{fig:3}, but for the isovector magnetic moments. Reproduced from Ref.~\cite{Li2011Sci.ChinaPhys.Mech.Astron.204}.}
\end{figure}

In Figures~\ref{fig:3} and~\ref{fig:4}, the relativistic results with considering MEC and second-order corrections are compared
with data. The results for CDFT represent the calculations with the time odd fields in Ref.~\cite{Li2011Sci.ChinaPhys.Mech.Astron.204}. In
Fig.~\ref{fig:3}, it is shown clearly that the mean-field calculations
have already provided an excellent description of the isoscalar
magnetic moments. As discussed in Ref.~\cite{Li2011Sci.ChinaPhys.Mech.Astron.204}, the effects of MEC on isoscalar magnetic moments
are negligible. The effects of 2nd have also only a very small
influence on the isoscalar magnetic moments of nuclei with $A=17$
and $A=41$, but they lead to relatively large corrections for
nuclei with $A=15$ and $A=39$. Such 2nd corrections enhance the
discrepancy between the calculated values and the data. It should be
pointed out that the operator $\hat{V}\hat{Q}/(E_n-\hat{H}_0)$ in
eq.~(\ref{eq:sec}) does not commute with the magnetic moment operator
$\hat{\mu}$ in eq.~(\ref{mm-operator}) and that this gives rise
to non-zero corrections for the isoscalar magnetic moment from
2nd.

In Fig.~\ref{fig:4}, it is shown that the mean-field calculations
produce values that are larger than the data for the particle
states with $A=17$ and $A=41$, and approximately equal to the data
for the hole states $A=15$ and $A=39$. Therefore the positive
contribution from the one-pion exchange currents worsen the
description of isovector magnetic moments. Here the second-order corrections gives a significant negative correction to the
isovector magnetic moments. This is consistent with the
non-relativistic calculations in Refs.~\cite{Towner1987Phys.Rep.263,Arima1987Adv.Nucl.Phys.1},
and finally improves the agreement with data, especially for $A=17$
and $A=41$. From calculations with other effective interactions
one comes to the same conclusions. As discussed in Ref.~\cite{Li2011Sci.ChinaPhys.Mech.Astron.204}, the isovector magnetic moments are relatively well
reproduced with the net effect
between the second-order and the one-pion exchange current
corrections, especially for $A=17$ and $A=41$.

On the whole, both the isoscalar and isovector magnetic moments of $LS$ closed shell nuclei plus or
minus one nucleon have been well explained by the CDFT with MEC and CM. The polarization effect from time odd fields, the one-pion
exchange current and the configuration mixing corrections are important and work together
to improve the description of the magnetic moments.

\section{Magnetic moments for nuclei with $jj$ closed shell core plus or minus one nucleon}

In Ref.~\cite{Furnstahl1987Nucl.Phys.A539}, magnetic moments for nuclei with $jj$ closed shell core near $^{208}$Pb have been studied in CDFT including the contribution from the core. The corresponding results show an improvement in comparison with the valence-nucleon approximation. On the other hand, meson exchange currents and configuration mixing, especially the first-order configuration mixing, are very important for the description of the magnetic moment of such nuclei.

In Ref.~\cite{Li2013Phys.Rev.C064307}, the magnetic moments in nuclei with $jj$ closed shell core plus or minus one nucleon, $^{207}$Pb, $^{209}$Pb, $^{207}$Tl, and $^{209}$Bi, are investigated in CDFT including time-odd fields, one-pion exchange currents, and first- and second-order configuration mixing corrections. In these calculations, the relativistic point-coupling model is adopted.

In Table.~\ref{tab1}, the first-order configuration mixing corrections to the magnetic moment of $^{209}$Bi are given. They are obtained
from relativistic calculations using the density functional PC-PK1~\cite{Zhao2010Phys.Rev.C054319} and also compared with the non-relativistic results using different interactions: Kallio-Kolltveit (KK)~\cite{Green1965Phys.Lett.142}, Gillet~\cite{Gillet1964Phys.Lett.44}, Kim-Rasmussen (KR)~\cite{Kim1963Nucl.Phys.184}, Brueckner~\cite{Brueckner1958PhysRev.109.1023}, Hamada-Johnston (HJ)~\cite{Hamada1962Nucl.Phys.382} potential, and the M3Y~\cite{Bertsch1977Nucl.Phys.A399} interaction. The corresponding first-order corrections are taken from Refs.~\cite{Mavromatis1966Nucl.Phys.545,Mavromatis1967Nucl.Phys.A17,Arima1972Phys.Lett.B435} respectively. For the relativistic calculations, both results with and without considering the residual interaction provided by pion exchange are also presented.

\begin{table*}
\tabcolsep=1.6pt
\centering
\caption{
\label{tab1}
First-order configuration mixing corrections to the magnetic moment of $^{209}$Bi obtained from relativistic calculations using the PC-PK1 effective interaction, in comparison with non-relativistic results using different interactions. In the relativistic calculations, both results with and without considering the residual interaction provided by pion are given. Reproduced from Ref.~\cite{Li2013Phys.Rev.C064307}}
\begin{ruledtabular}
\begin{tabular}{ccrcrcccccrc}
  & \multicolumn{8}{c}{Non-rel.} && \multicolumn{2}{c}{Rel.}\\ \cline{2-9}\cline{11-12}
Interactions & KK & Gillet & KR. I & KR. II & Brueckner & HJ & ~Kuo~ & M3Y && \multicolumn{2}{c}{PC-PK1\cite{Li2013Phys.Rev.C064307}}
\\ \cline{3-7}\cline{11-12}
Ref.&\cite{Mavromatis1966Nucl.Phys.545}&\multicolumn{5}{c}{\cite{Blomqvist1965Phys.Lett.47,Mavromatis1967Nucl.Phys.A17}}&\cite{Arima1972Phys.Lett.B435}&\cite{Bertsch1977Nucl.Phys.A399} && \multicolumn{1}{c}{without $\pi$}&\multicolumn{1}{c}{with $\pi$}\\\hline
${\pi}(1h_{{9}/{2}}1h_{{11}/{2}}^{-1})$
& 0.37 & 0.46 & 0.53 & 0.70 & 0.71 & 0.55 &  &0.43 &~~~&$-$0.11&0.19\\
${\nu}(1i_{{11}/{2}}1i_{{13}/{2}}^{-1})$
& 0.15 &$-$0.02 & 0.00 & $-$0.06 & 0.04 & 0.25 & &0.24 && 0.07 & 0.33\\
Total  & 0.52 & 0.43 & 0.53 & 0.64 & 0.75 & 0.80 & 0.79 &0.68 && $-0.04$ & $0.52$\\
\end{tabular}
\end{ruledtabular}
\end{table*}

In non-relativistic calculations, only the excitations of spin-orbit partners can contribute to the first-order magnetic moment correction, because of selection rules imposed by the magnetic moment single-particle operator. In relativistic calculations this selection rule is only approximately valid. However, in the present relativistic calculations, only two particle-hole excitations can contribute to the first-order magnetic moment correction of $^{209}$Bi, i.e., the ${\pi}(1h_{{9}/{2}}1h_{{11}/{2}}^{-1})$ and ${\nu}(1i_{{11}/{2}}1i_{{13}/{2}}^{-1})$ excitations, and all other particle-hole excitations give a small and negligible contribution to its expectation value in first-order perturbation theory.

 \begin{figure}[!ht]
 \centering
 \includegraphics[width=8.5cm]{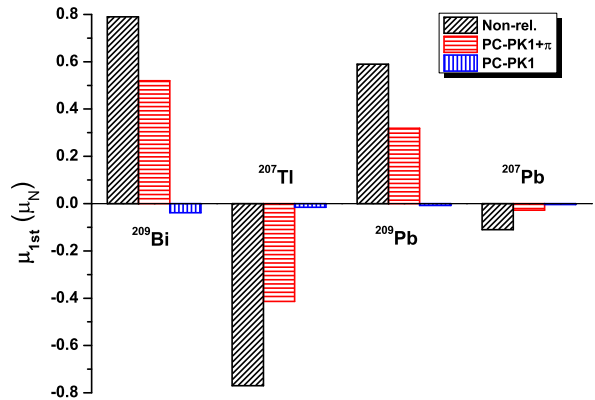}\vspace*{-0.5cm}
  \caption{First-order configuration mixing corrections to magnetic moments of $^{209}$Bi, $^{207}$Tl, $^{209}$Pb and $^{207}$Pb obtained from relativistic calculations using PC-PK1 interaction, in comparison with the non-relativistic results obtained from Ref.~\cite{Arima1972Phys.Lett.B435}. In the relativistic calculations, the results with and without $\pi$ are given. Reproduced from Ref.~\cite{Li2013Phys.Rev.C064307}}
 \label{jj_fig5}
 \end{figure}

As shown in Table~\ref{tab1}, the non-relativistic calculations give remarkable first-order corrections ($0.43\,\mu_N\sim0.80\,\mu_N$),
while the corresponding corrections given by relativistic calculations using the PC-PK1 effective interaction are very small ($-0.03\,\mu_N$)
and can be neglected. Only after the residual interaction provided by the pion is included, PC-PK1 gives significant corrections ($0.59\,\mu_N$)
that are consistent with non-relativistic results.

The first-order configuration mixing corrections to the magnetic moments of $^{207}$Pb, $^{209}$Pb, $^{207}$Tl and $^{209}$Bi obtained from relativistic calculations using the PC-PK1 interaction are shown in Fig.~\ref{jj_fig5} and compared with non-relativistic results obtained from Ref.~\cite{Arima1972Phys.Lett.B435}, in order to further confirm the effects of the residual interaction provided by pion. It is easy to see that without the residual interaction provided by pion, the relativistic calculations give negligible first-order corrections to the magnetic moments of all four nuclei. If the residual interaction provided by pion is included, relativistic calculations are in reasonable agreement with non-relativistic results for all present nuclei.

As seen in Eq.~(\ref{eq:1cp-bq}) for the correction to the magnetic moment, the excitation energy, and the interactions can all lead to differences between relativistic and non-relativistic results. It is well known that the effective mass is relatively small in self-consistent calculations based on density functional theory, which leads to an increased gap at the Fermi surface in the single particle spectrum and to larger $ph$-energies. Taking into account the energy dependence of the self energy in the framework of couplings to low-lying collective surface modes, considerably larger effective masses and smaller energy gaps have been found in the literature~\cite{Ring1973Nucl.Phys.A198,Bernard1980Nucl.Phys.A75,Litvinova2006Phys.Rev.C044328,Ring2009Phys.At.Nucl.1285},
which are closer to the experimental values. As discussed in Ref.~\cite{Li2013Phys.Rev.C064307}, the experimental single particle energies are used rather than the self-consistent CDFT single particle energies in the intermediate states for a relativistic estimation.
Since the non-relativistic results are obtained with experimental energy splittings, it is found that by adopting experimental excitation
energy, the relativistic calculations give almost
the same first-order corrections as the results by adopting self-consistent CDFT single particle energies shown in Fig.~\ref{jj_fig5}.
Although there is some difference between the matrix elements of the relativistic and the non-relativistic magnetic moment operator, the difference in the first-order corrections is subtle as shown in Fig.~\ref{jj_fig5}. Therefore, the difference between relativistic and non-relativistic results are mainly due to interactions. The
residual interaction provided by pion plays an important role in the relativistic descriptions of nuclear magnetic moments and it has been included in the following calculations of second-order corrections.

\begin{figure*}[!ht]
\centering
\includegraphics[width=8.5cm]{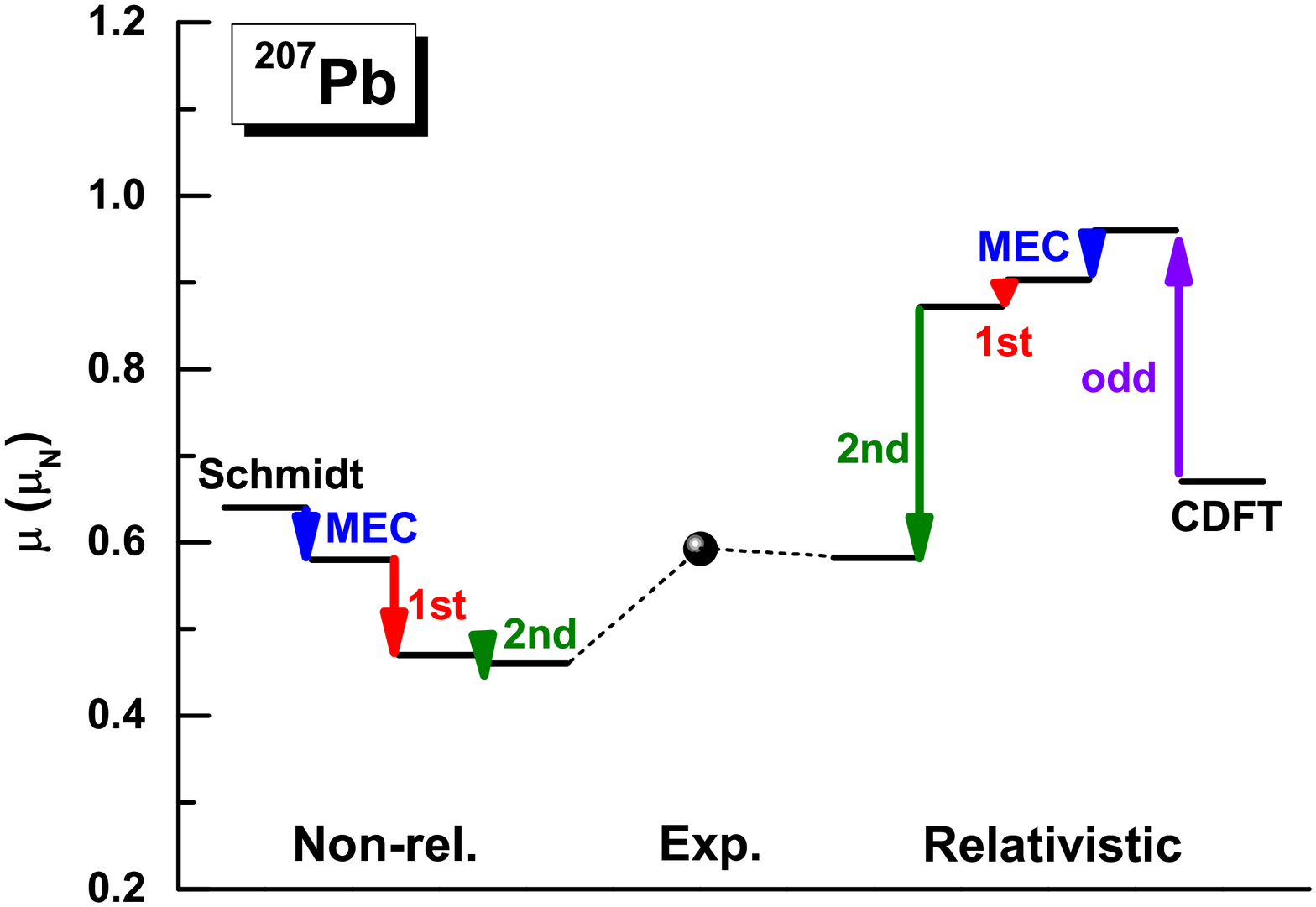} \hspace*{-0.8cm}
\includegraphics[width=8.5cm]{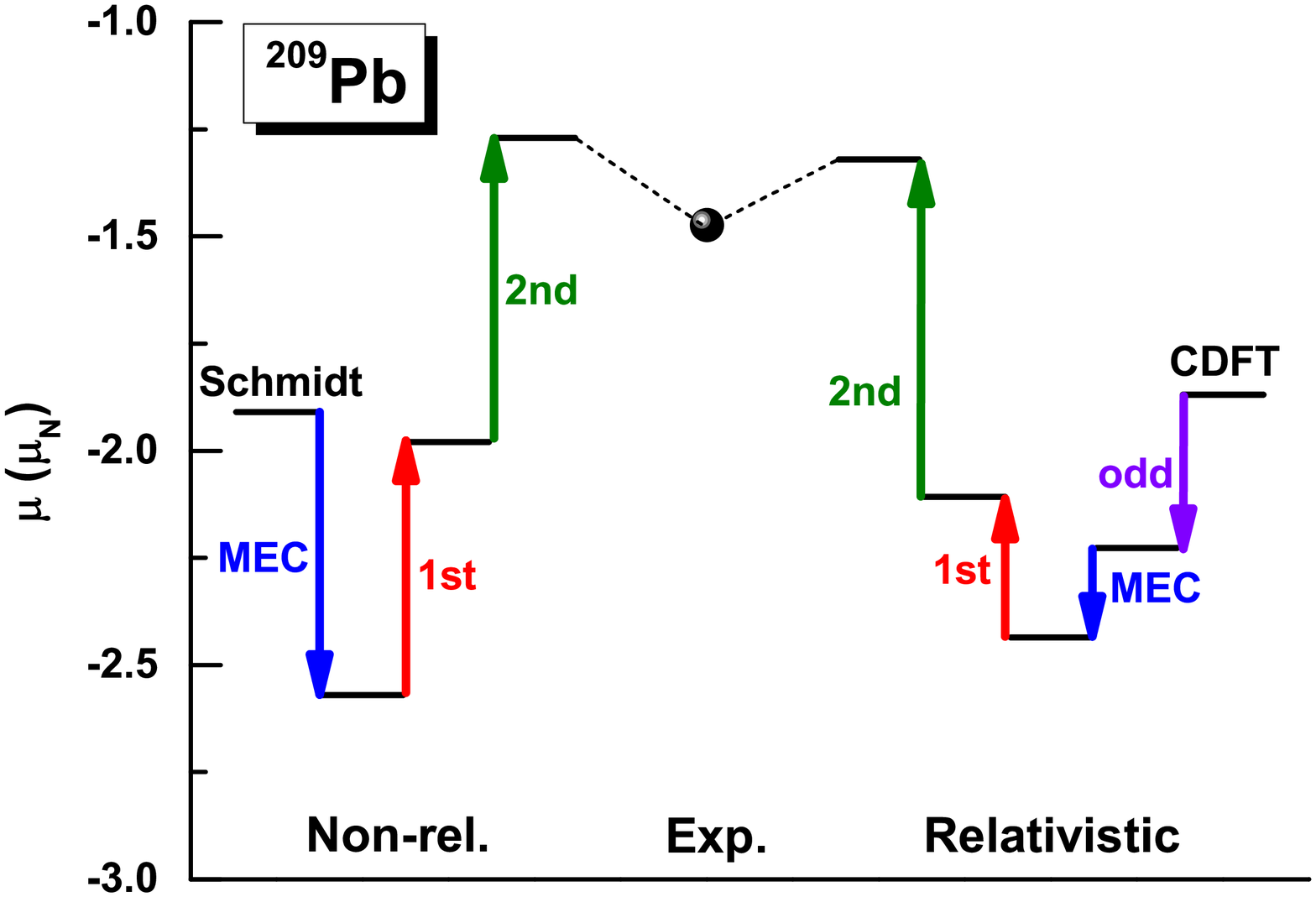} \vspace*{-1.0cm}\\
\includegraphics[width=8.5cm]{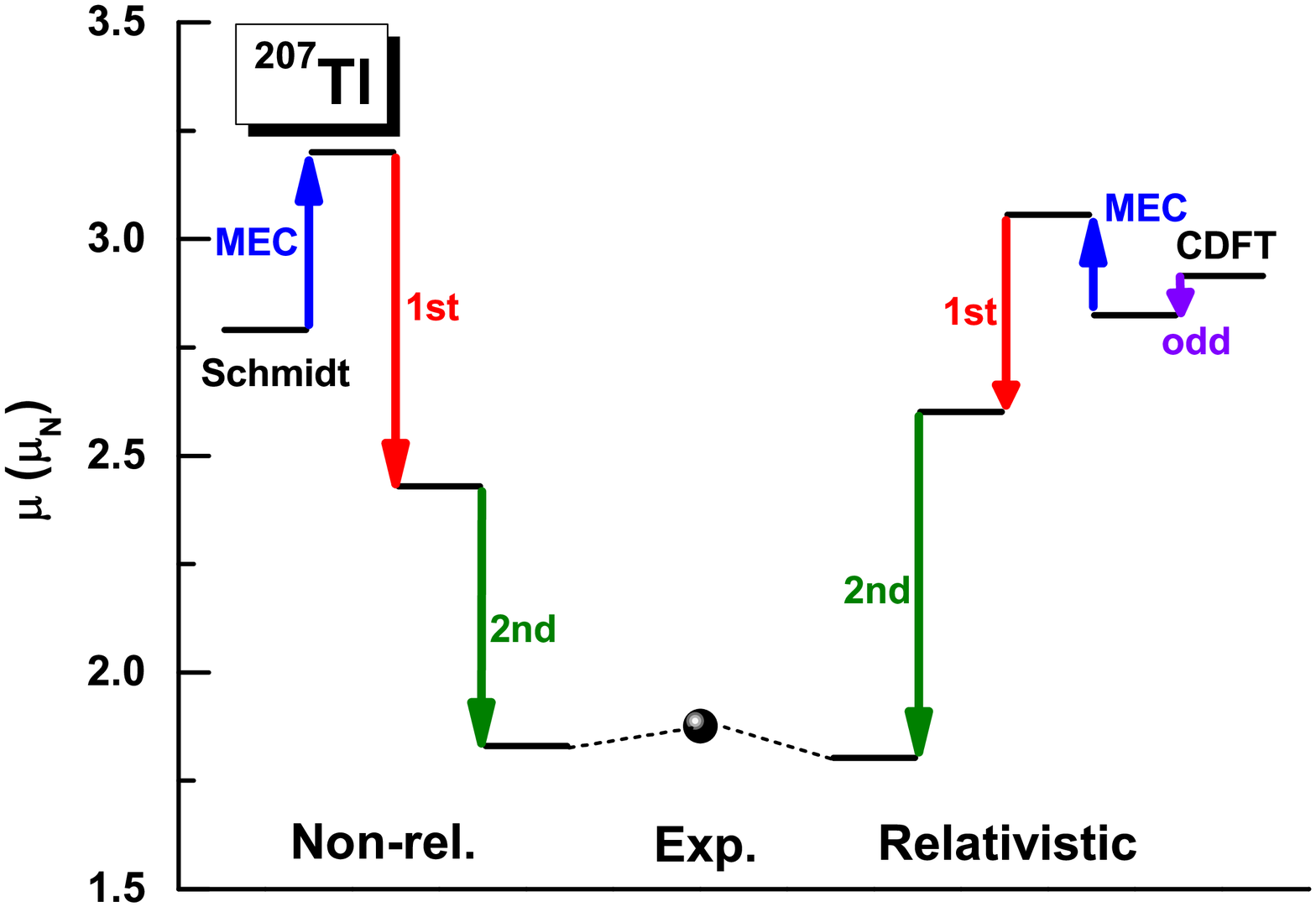} \hspace*{-0.8cm}
\includegraphics[width=8.5cm]{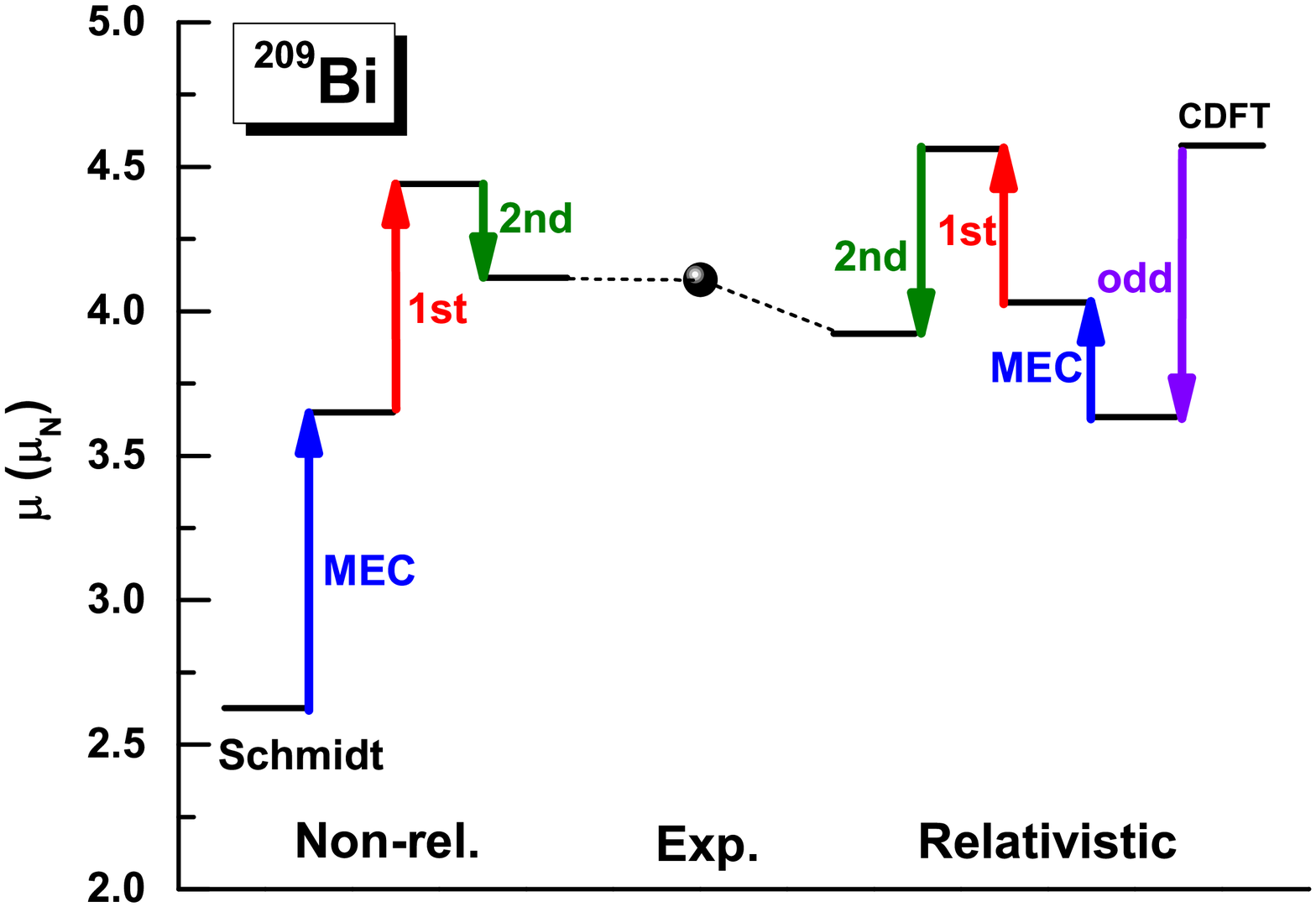} \vspace*{-0.5cm}
\caption{Magnetic moments of the nuclei $^{207}$Pb, $^{209}$Pb, $^{207}$Tl and $^{209}$Bi, obtained from relativistic calculations using the PC-PK1 interaction with time odd contribution, meson exchange currents, first- and second-order corrections, in comparison with data (solid circle) and the corresponding non-relativistic results from Ref.~\cite{Arima1987Adv.Nucl.Phys.1}. Reproduced from Ref.~\cite{Li2013Phys.Rev.C064307}}
\label{jj_fig7}
\end{figure*}

Fig.~\ref{jj_fig7} shows the final results for the magnetic moments of $^{207}$Pb, $^{209}$Pb, $^{207}$Tl and $^{209}$Bi. They are obtained from CDFT with PC-PK1 and corresponding corrections are added: time-odd fields (labeled as odd), meson exchange currents (labeled as MEC), first- (labeled as 1st) and second-order (label as 2nd) configuration mixing. These relativistic results are compared with data (labeled as solid circle) and non-relativistic results from Ref.~\cite{Arima1987Adv.Nucl.Phys.1}. The magnetic moments obtained from spherical CDFT are labeled as CDFT. The differences between magnetic moments of CDFT with time-odd fields and magnetic moments of spherical CDFT represent the corrections due to time-odd fields. In the relativistic calculations, MEC only contains the one-pion exchange current correction, while the MEC in non-relativistic calculations includes the one-pion exchange current, the $\Delta$ isobar current and the crossing term between MEC and first-order configuration mixing. For first- and second-order corrections in the relativistic calculation, the residual interaction provided by the poin is included.

As discussed in Ref.~\cite{Li2013Phys.Rev.C064307}, the magnetic moments of all four nuclei in the relativistic calculations are considerably improved by including first-order corrections,
MEC and second-order corrections, and they are now in agreement with non-relativistic results. The magnetic moment of $^{207}$Pb is excellently reproduced by relativistic calculations, while the corresponding deviation from data is $0.01\,\mu_N$, and much better than the non-relativistic deviation $0.05\,\mu_N$.
For $^{209}$Pb, the deviations from data are about $0.15\,\mu_N$ and $0.2\,\mu_N$ respectively for relativistic and non-relativistic results. For $^{207}$Tl, both the non-relativistic description and the relativistic description are very good, as the corresponding two deviations from data are less than $0.1\,\mu_N$.
The magnetic moment of $^{209}$Bi is also well reproduced by relativistic and non-relativistic calculations, and the relative deviations from data for both calculations are less than $5\%$. On the whole, the relative deviation  of the present four nuclei in the relativistic calculation is $6.1\%$, better than the corresponding non-relativistic results $13.2\%$~\cite{Arima1987Adv.Nucl.Phys.1}, as shown in Ref.~\cite{Li2013Phys.Rev.C064307}.

It is obvious that the first-order, MEC and second-order corrections given by relativistic calculations have the same sign and order of magnitude as the corresponding corrections given by non-relativistic calculations. This further shows that the present relativistic calculations are reasonable, including the appropriate treatment of truncation in second-order corrections.

\section{Magnetic moments of deformed odd-$A$ nuclei}

The odd-A nuclei with either $LS$ and $jj$ closed shell plus or minus one nucleon are usually spherical. Therefore the above discussion is mainly for the nucleus with spherical shape. In the nuclear chart, except those near the doubly closed shell, most nuclei are deformed. Here we will discuss the description of the magnetic moments for deformed odd-$A$ nuclei in CDFT.

In particular, apart from the magnetic moments of stable nuclei~\cite{Stone2014},
it is now even possible to measure the nuclear magnetic moments of
many short-lived nuclei far from the stability line with high
precision~\cite{Neyens2003Rep.Prog.Phys.633} with the development of the radioactive
ion beam  technique.

For deformed odd-$A$ nuclei, the valence nucleon approximation is invalid as there is a strong coupling between the core and the valence nucleon, which can not be treated in the perturbation theory. Finally, the total magnetic moment consists of two parts, i.e., the intrinsic nucleonic motion and the collective rotational motion~\cite{Bohr1975}. Thus, it is necessary to study the ground-state magnetic moment of deformed odd-$A$ nuclei in deformed CDFT as a first step.

As discussed in Ref.~\cite{Li2009Sci.ChinaSer.G1586}, the nuclear magnetic moments of $^{33}$Mg has become a hot
topic due to the following reasons: 1) it is a neutron-rich nucleus
close to the so-called ``island of inversion"~\cite{Warburton1990Phys.Rev.C1147} proposed as the
unusual features for a group of neutron-rich nuclei in a region of the nuclear chart far from stability line;
2) different spins and configurations for the ground state of
$^{33}$Mg are assigned in a series of
experiments~\cite{Nummela2001Phys.Rev.C054313,Pritychenko2002Phys.Rev.C061304,Elekes2006Phys.Rev.C044314,Tripathi2008Phys.Rev.Lett.142504}.
In order to remove the confusion, the spin and magnetic moment for
the ground state in $^{33}$Mg have been directly measured in
Ref.~\cite{Yordanov2007Phys.Rev.Lett.212501} with $I=3/2$ and
$\mu=-0.7456(5)\mu_\mathrm{N}$, which becomes a test for various
theoretical approaches. In shell-model, the magnetic moment of the
ground-state in $^{33}$Mg, can be reproduced only in the model space
with $2p-2h$ configuration~\cite{Yordanov2007Phys.Rev.Lett.212501}. With the assignment
of configuration in Ref.~\cite{Tripathi2008Phys.Rev.Lett.142504}, the simple Additivity
Rules~\cite{Lawson1980} can only account for half of the
experimental magnetic moments.

\begin{figure}[!ht]
\centering
\includegraphics[width=8cm]{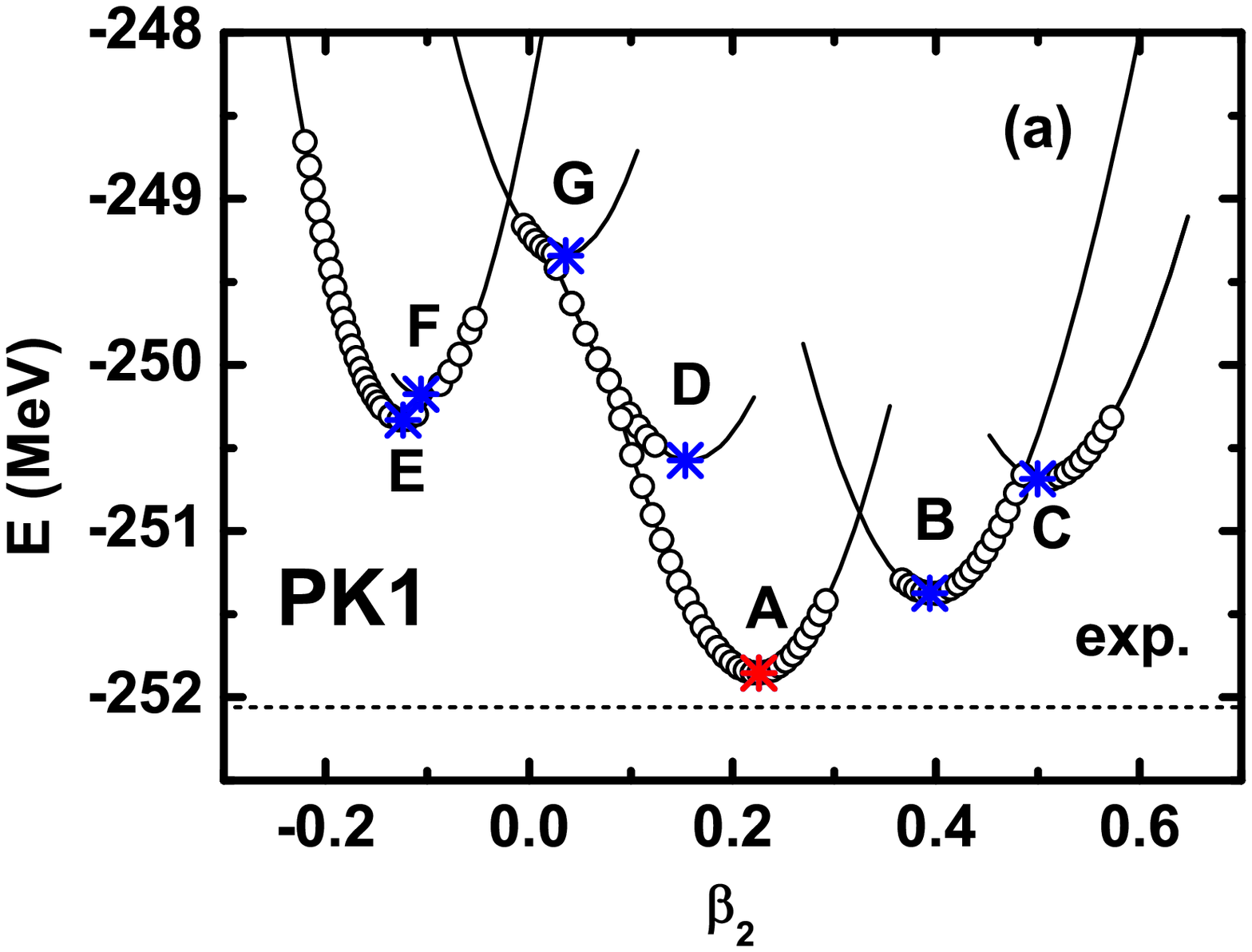}\\ \vspace*{-0.6cm}
\includegraphics[width=8cm]{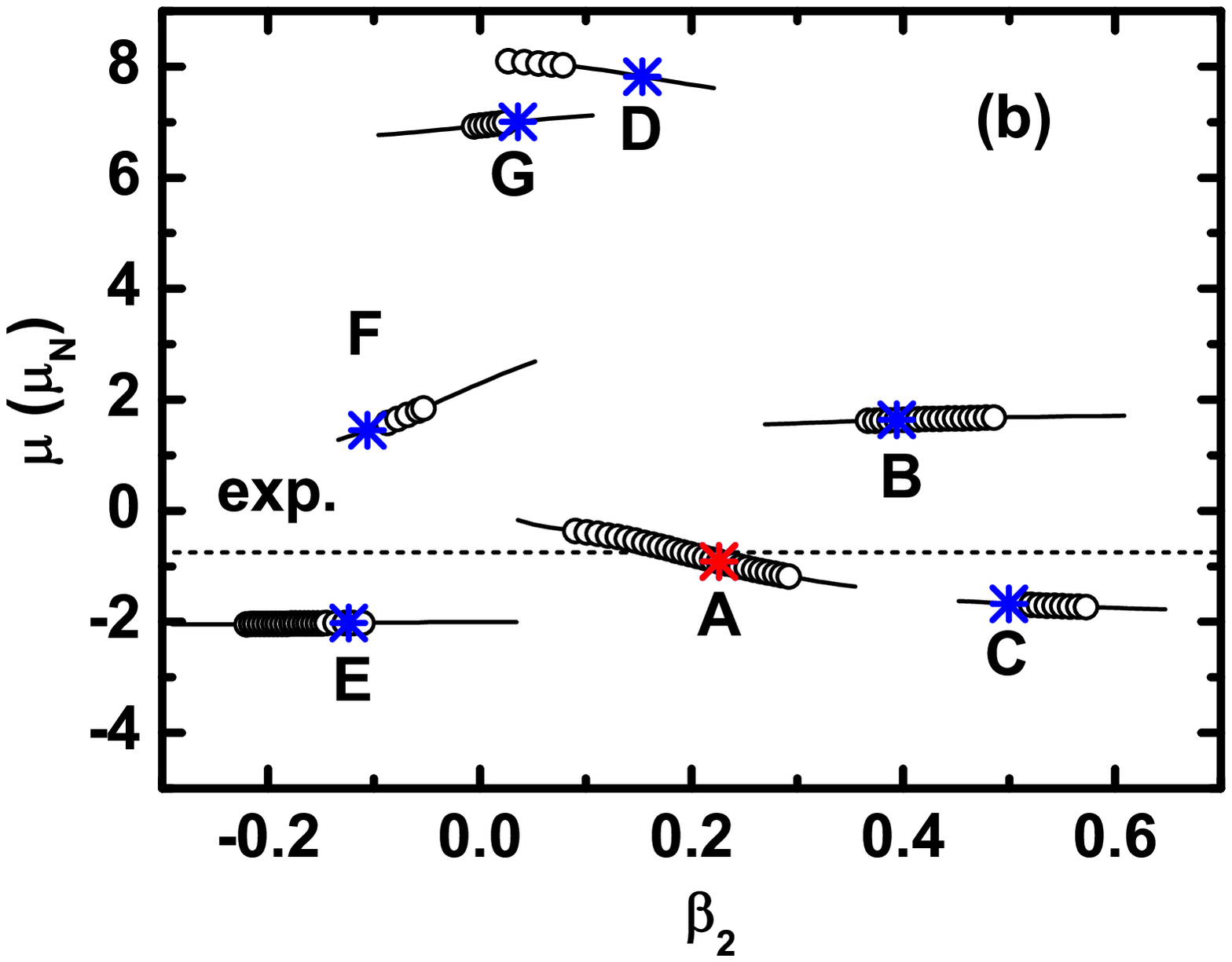}
\caption{\label{Mg33} (a) The energy surfaces for
$^{33}$Mg as a function of $\beta_2$ by adiabatic (open circles) and
configuration-fixed (solid lines) deformation constrained CDFT
approach with time-odd component using PK1 parameter set. The minima
in the energy surfaces for fixed configurations are represented as
stars and respectively, labeled as A, B, C, D, E, F, and G. (b) Magnetic moments for the
corresponding configurations in panel (a) as a function of
$\beta_2$. Reproduced from Ref.~\cite{Li2009Sci.ChinaSer.G1586}.}
\end{figure}

\begin{figure}[!ht]
\centering
\includegraphics[width=8cm]{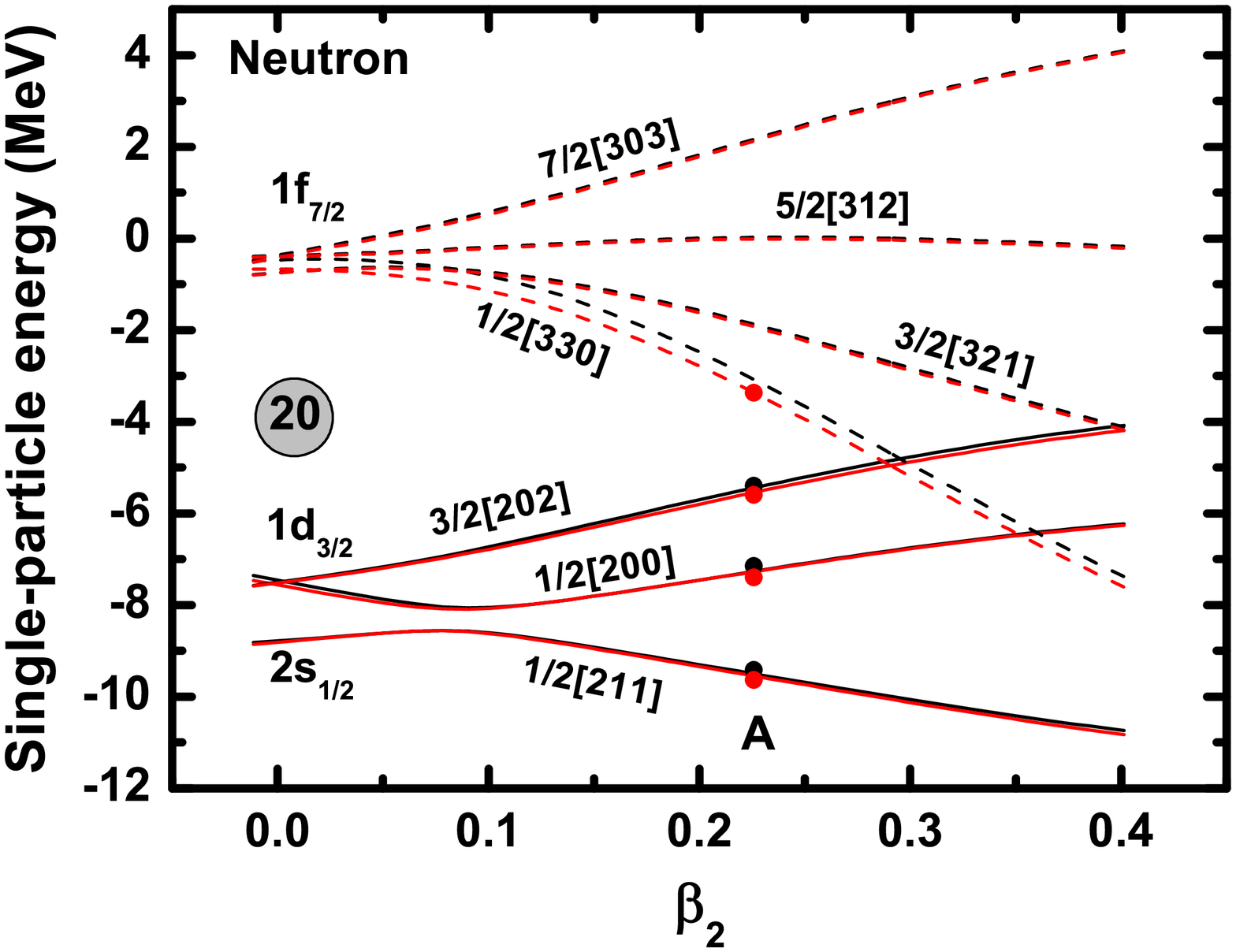}
\caption{\label{Mg33_level} Neutron single-particle
energies for $^{33}$Mg as a function of $\beta_2$ obtained by
configuration-fixed deformation constrained calculation for the
configuration of ground state A. Positive (negative) parity states
are marked by solid (dashed) lines. Each pair of time reversal
conjugate states splits up into two levels with the third component
of total angular momentum $\Omega>0$ and $\bar{\Omega}<0$ denoted by
red and black lines respectively. The solid circle denotes that the
corresponding orbitals are occupied in the ground state. Reproduced from Ref.~\cite{Li2009Sci.ChinaSer.G1586}.}
\end{figure}

The energy surfaces for $^{33}$Mg as a function of the quadrupole
deformation parameter $\beta_2$ calculated by adiabatic, shown as
open circles, and configuration-fixed, shown as solid lines,
deformation constrained CDFT approach with time-odd component using
PK1~\cite{Long2004Phys.Rev.C034319} are presented in Fig.~\ref{Mg33}(a). As discussed
in the Ref.~\cite{Lu2007Eur.Phys.J.A273}, the configuration-fixed deformation
constrained calculation gives a continuous and smooth curve for the
energy surfaces as a function of $\beta_2$. The local minima in the
energy surfaces for each configuration are represented by stars and
labeled as A¨CG in ascending order of energy. The ground state A is
found to be prolate deformed, $\beta_2=0.23$, with the total energy
$ -251.85$ MeV, which is close to the data $-252.06$
MeV~\cite{Wang2017Chin.Phys.C030003}. Using Eq.~(\ref{magnetic-moment}), the
effective electromagnetic current gives the nuclear magnetic moment
for different configurations in Fig.~\ref{Mg33}(b). It is found that the
magnetic moment is sensitive to the configuration, but not so much
to $\beta_2$. The magnetic moment of the ground state in CDFT is $
-\,0.913\,\mu_N$ which is in good agreement with the data $ \mu =
-\,0.7456(5)\,\mu_N$~\cite{Yordanov2007Phys.Rev.Lett.212501}, compared with the
shell-model results $-\,0.675\,\mu_N$ and
$-\,0.705\,\mu_N$~\cite{Yordanov2007Phys.Rev.Lett.212501} restricted to $2p$-$2h$
configuration using two different interactions designed specifically
for the island of inversion.

In Fig.~\ref{Mg33_level}, in order to examine the evolution of the single particle level and
compare with the results in Ref.~\cite{Yordanov2007Phys.Rev.Lett.212501}, neutron
single-particle energies for $^{33}$Mg as a function of $\beta_2$
obtained by configuration-fixed deformation constrained calculation
for the configuration of ground state A are presented. The positive (negative) parity states are marked by
solid (dashed) lines, and the occupied orbitals are represented by
filled circles. As discussed in Ref.~\cite{Li2009Sci.ChinaSer.G1586}, the self-consistent calculation gives the odd
neutron in $1/2[330]$ orbital with $\beta_2=0.23$ for the ground
state, while in Ref.~\cite{Yordanov2007Phys.Rev.Lett.212501}, in a Nilsson-model
picture, the odd neutron in $3/2[321]$ orbital with prolate
deformation $0.3 < \beta_2 < 0.5$ is proposed to reproduce the spin
and parity $I^\pi=\frac{3}{2}^-$.

\section{Summary}

In summary, the studies on nuclear magnetic moments in CDFT has been reviewed. By considering the time-odd fields, one-pion exchange current, first-order, and second-order corrections, the magnetic moments of odd-$A$ nuclei with $LS$ and $jj$ closed shell plus or minus one nucleon have been reproduced and this method should be extended to apply for more nuclei with a doubly closed shell plus or minus one nucleon such as $^{133}$Sb~\cite{Stone1997Phys.Rev.Lett.820}, $^{67}$Ni and $^{69}$Cu~\cite{Rikovska2000Phys.Rev.Lett.1392}, and $^{49}$Sc~\cite{Ohtsubo2012Phys.Rev.Lett.032504}. In addition, the descriptions of the magnetic moments in deformed odd-$A$ nuclei become possible. It should be noted that the descriptions of nuclear magnetic moments with other models such as shell model~\cite{Caurier2005Rev.Mod.Phys.427,Zhao2014Phys.Rep.1,Bian2007Phys.Rev.C014312} have not been discussed.
The magnetic dipole moments of most atomic nuclei throughout the periodic table still remain unexplained and the underlying physics mechanism is not fully understood. We are looking forward to more contributions to this important subject in the future.

There are still many important open questions for the descriptions of nuclear magnetic moments in the CDFT. So far, the contributions of the Dirac sea have not been included in the configuration mixing calculations, because they are far from the configurations space under consideration. The crossing terms between MEC and configuration mixing as well as the influence of higher order diagrams in RPA type configuration mixing calculations are also neglected~\cite{Bauer1973Nucl.Phys.A535}. An additional point not included so far is the coupling to the $\Delta$ isobar current~\cite{Rho1974Nucl.Phys.A535,Oset1979Phys.Rev.Lett.47,Knupfer1980Phys.Lett.B349}.
Of course, it will be also interesting to study the influence of other successful covariant density functionals on the market, in particular those bases on relativistic Hartree-Fock theory~\cite{Long2006Phys.Lett.B150}, where the pion and the resulting tensor forces can be included in a self-consistent way.

\begin{acknowledgments}
This work is dedicated to Prof. Akito Arima on the occasion of his 88 birthday, i.e., rice anniversary in Chinese character.
Prof. Akito Arima has made important contributions not only in nuclear physics but also in promoting nuclear physics
research and collaboration worldwide. It is a great honor that we have the opportunity to collaborate with Prof. Akito Arima  on pseudospin symmetry and magnetic moments. The successful collaborations have deepened our understanding of the related fields.

We would like to thank Akito Arima, Jinniu Hu, Haozhao Liang, Peter Ring, Jixuan Wei, Jiangming Yao and Ying Zhang for discussions and collaborations. This work is partly supported by the National Key R\&D Program of China (2018YFA0404400), Natural Science Foundation of China (Grants No. 11335002, No. 11621131001, and No. 11675063), and China Scholarship Council (No. 201706175122).
\end{acknowledgments}

\end{document}